\documentclass[authoryear,a4paper,11pt]{elsarticle} 
\journal{EJOR}

\pdfoutput=1 

\usepackage{amsmath,amssymb,natbib,graphicx,hyperref,setspace,enumitem,soul} 
\usepackage[left=1in,right=1in,top=1.4in,bottom=1.4in]{geometry}

\hypersetup{
	pdftitle={Simulation schemes for the Heston model with Poisson conditioning},
	pdfauthor={Jaehyuk Choi and Yue Kuen Kwok},
	colorlinks=true,
	linkcolor=blue,
	citecolor=blue,
	urlcolor=blue,
	bookmarksnumbered=true,
	pdfstartview=
}

\newcommand{\var}{V}
\newcommand{\ivt}{I_{0,T}}
\newcommand{\ivtm}{I_{0,T}^m}
\newcommand{\ivi}{I_{0,N}}
\newcommand{\ivstep}{I_{i,i+1}}
\newcommand{\vov}{\xi}
\newcommand{\mr}{\kappa}
\newcommand{\vinf}{\theta}
\newcommand{\dt}{h}

\newcommand{\qtext}[2][\quad]{#1\text{#2}#1}

\newcommand{\GAM}{\Gamma_{\!1}}
\newcommand{\NCX}{\chi^2}
\newcommand{\POIS}{\text{POIS}}
\newcommand{\BES}{\text{BES}}
\newcommand{\IG}{\text{IG}}

\begin{document}
\begin{frontmatter}

\title{Simulation schemes for the Heston model with Poisson conditioning}

\author[phbs]{Jaehyuk Choi} \corref{corrauthor}
\ead{jaehyuk@phbs.pku.edu.cn}
\author[hkust]{Yue Kuen Kwok}
\ead{maykwok@hkust-gz.edu.cn}

\cortext[corrauthor]{Correspondence. \textit{Tel:} +86-755-2603-0568, \textit{Address:} Rm 755, Peking University HSBC Business School, University Town, Nanshan, Shenzhen 518055, China}
\address[phbs]{Peking University HSBC Business School, University Town, Nanshan, Shenzhen 518055, China}
\address[hkust]{Financial Technology Thrust, Hong Kong University of Science and Technology, Guangzhou 511453, China}

\date{October 31, 2023}

\begin{abstract}
Exact simulation schemes under the Heston stochastic volatility model (e.g., Broadie--Kaya and Glasserman--Kim) suffer from computationally expensive modified Bessel function evaluations. We propose a new exact simulation scheme without the modified Bessel function, based on the observation that the conditional integrated variance can be simplified when conditioned by the Poisson variate used for simulating the terminal variance. Our approach also enhances the low-bias and time discretization schemes, which are suitable for pricing derivatives with frequent monitoring. Extensive numerical tests reveal the good performance of the new simulation schemes in terms of accuracy, efficiency, and reliability when compared with existing methods.
\end{abstract}
\end{frontmatter}

\section{Introduction} \noindent
The \citet{heston1993closed} stochastic volatility model relaxes the constant volatility assumption in the Black--Scholes (BS) model by taking the instantaneous variance to follow the square root diffusion process with mean reversion, commonly called the Cox--Ingersoll--Ross~(CIR, \citet{cox1985cir}) process. It is the most popular stochastic volatility model for market practitioners because of its analytic tractability in computing the prices of European options. Recently, the multifactor versions of the model have received attention as they can better explain the stylized facts observed in the derivative market~\citep{dafonseca2008multifactor,christoffersen2009shape,trolle2009generala,jaber2019lifting}. Owing to the existence of a closed-form formula for the characteristic function of the log-asset price, model calibration of market-observable European option prices can be performed efficiently using the Fourier inversion algorithm~\citep{lewis2000sv,lord2010complex}. For pricing path-dependent options under the Heston model, Monte Carlo (MC) simulations are often used. However, the standard Euler and Milstein time discretization simulation schemes suffer from a high bias owing to the square root of the diffusion function in the variance process. Negative values of variance from the simulation must be heuristically set to zero before taking the square root of variance. In addition, the square root function violates the Lipschitz condition; therefore, the convergence properties of the discretization scheme may not be guaranteed. There have been numerous fixes to these issues to minimize the discretization biases. A comprehensive review of these discretization schemes using various fixes can be found in \citet{lord2010comparison}.

A major breakthrough was made by \citet{broadie2006mcheston} in the simulation of the Heston model from its exact distribution. Their ``exact'' simulation\footnote{The term ``exact'' simulation was coined by \citet{broadie2006mcheston}. Despite the name, the overall simulation algorithm still requires numerical approximations such as the use of discrete Fourier inversion algorithm. They used the term ``exact'' to distinguish their algorithm from the usual Euler discretized schemes that involve discretization bias. The term ``exact'' has gained wide acceptance in the literature as the norm in several dozen papers that deal with simulation methods. Therefore, we also follow the common norm of ``exact''.} procedures consist of three steps: (i) sampling of the terminal variance conditional on the initial variance, (ii) sampling the integrated variance conditional on the initial and terminal variance values, and (iii) sampling the asset price process conditional on the variance and integrated variance. As the exact simulation approach avoids simulation bias, simulation errors remain inversely proportional to the square root of the computational time budget. However, the Broadie--Kaya exact simulation algorithm is not competitive in accuracy-speed comparison since it requires extensive computational time to sample the conditional integrated variance via the numerical inversion of the Laplace transform in each simulation path. To improve computational efficiency, one may use a caching technique to sample the terminal variance and conditional integrated variance via precomputation and interpolation of the appropriate inverse distribution functions~\citep{vanhaastrecht2010efficient,zeng2023analytical}. 
Despite its limitations, \citet{broadie2006mcheston}'s pioneering work triggers the construction of exact simulation schemes for other stochastic volatility models, such as the stochastic-alpha-beta-rho (SABR)~\citep{cai2017sabr,choi2019nsvh}, 3/2~\citep{baldeaux2012exact,zeng2023analytical}, Wishart~\citep{kang2017exact},
and Ornstein--Uhlenbeck-driven stochastic volatility model~\citep{li2019exact_ou,choi2023ousv}.

Based on \citet{pitman1982decom}'s decomposition of Bessel bridges, \citet{glasserman2011gamma} show that conditional integrated variance can be expressed as gamma expansions (GE). Significant computational speedup can be achieved by sampling the conditional integrated variance via the sums of the mixtures of gamma random variates (with an approximation of the truncated terms). In a related study, \citet{malham2021series} construct a new series expansion for the conditional integrated variance in terms of double infinite weighted sums of independent random variables through a measure change and decomposition of squared Bessel bridges. 

There have been various efforts to improve the time discretization schemes. \citet{tse2013lowbias} propose a low-bias simulation algorithm by approximating the conditional integrated variance with an Inverse Gaussian (IG) variate via matching mean and variance. With its lower computational cost per time step, the IG scheme can be used as a multiperiod scheme for pricing path-dependent options. \citet{andersen2008simple} constructs a time discretization scheme, where the variance at discrete time points is simulated via the quadratic-exponential (QE) approximation with martingale correction. However, he only uses the trapezoidal rule to approximate the conditional integrated variance. Such an approximation is acceptable when the time step is small. 

There is no one-size-fits-all solution among the simulation approaches. Each simulation method has its own advantages depending on the monitoring frequency of the derivatives to price. \citet{tse2013lowbias} conclude that the decision among GE, IG, or QE schemes depends on the compromise between computational cost and bias. The exact GE scheme is widely accepted as the best choice for European-style derivatives. Despite the high computational cost per step, we simply need to simulate just one step up to expiry. For path-dependent derivatives with frequent monitoring, the time-discretized QE scheme may be a better choice owing to its low computational cost per step. When the number of monitoring instants is moderate, one may choose to use the low-bias IG scheme as a compromise between exact and time discretization schemes.

Despite the advances in the Heston simulation algorithms, the computational efficiency still has room for improvement. The simulation methods mentioned above, except for the QE scheme, involve computationally demanding evaluations of the modified Bessel function. This has been criticized as a bottleneck in computation. In particular, the GE scheme~\citep{glasserman2011gamma} involves the Bessel random variable, the sampling of which takes up a significant portion of the computation time for the same reason. 

We propose enhanced Heston simulation schemes in all ranges based on the key observation that the conditional integrated variance can be further simplified when conditioned by the Poisson variate involved in the terminal variance. Consequently, the computationally trivial Poisson variate replaces the Bessel variate in the GE scheme~\citep{glasserman2011gamma}. In addition, by adopting the IG approximation~\citep{tse2013lowbias} for series truncation, the Poisson-conditioned GE scheme significantly enhances both speed and accuracy. The IG scheme~\citep{tse2013lowbias} is a special case of the new method, but the enhanced IG scheme no longer requires the modified Bessel function for the mean and variance calculations. \citet{broadie2006mcheston}'s Laplace inversion scheme can also benefit from our approach since Poisson conditioning removes the modified Bessel function from the Laplace transform. We also propose a Poisson-conditioned time discretization method with the corresponding martingale correction method, which is highly efficient for pricing derivatives with frequent monitoring. Our new time discretization scheme compares favorably with \citet{andersen2008simple}'s QE scheme.

The contributions of this study are summarized as follows. 
\begin{itemize}
	\item We propose the Poisson-conditioned GE scheme, an enhanced exact simulation scheme that achieves significant computational speedup by simplifying the conditional integrated variance via Poisson conditioning, thus avoiding the simulation of the tedious Bessel random variable required in earlier GE schemes.
	\item We improve the accuracy of the Poisson-conditioned GE by using the IG approximation~\citep{tse2013lowbias} in the series truncation procedure.
	\item We propose a Poisson-conditioned time discretization scheme for pricing discretely monitored path-dependent derivatives.
	\item Our enhanced schemes can flexibly be applied to the multifactor Heston model when the variance factors are independent with each other.
\end{itemize}

The remainder of this paper is organized as follows. In Section~\ref{s:model}, we introduce the Heston model and its analytical properties. We then review existing simulation algorithms and discuss the intrinsic computational challenges. In Section~\ref{s:pois}, we show how the Poisson conditioning can enhance existing algorithms. In Section~\ref{s:num}, we present extensive numerical tests that compare the performance of the Poisson-conditioned simulation schemes with existing simulation schemes for pricing European options and discretely monitored variance swaps. In Section~\ref{s:multi}, we extend our schemes to a class of the multifactor Heston models. Finally, we conclude this paper in Section~\ref{s:con}.

\section{Heston stochastic volatility model and existing simulation schemes} \label{s:model}
\noindent
The dynamics of the asset price process $S_t$ and its instantaneous variance process $\var_t$ of the Heston stochastic volatility model under a risk-neutral measure $Q$ are governed by the following coupled stochastic differential equations:
\begin{align}
\frac{\text{d} S_t}{S_t} &= (r-q) \;\text{d}t + \sqrt{\var_t} \left(\rho \;\text{d} Z_t + \sqrt{1-\rho^2} \;\text{d} W_t \right), \label{eq:hestona}\\
\text{d} \var_t &= \mr (\vinf - \var_t) \;\text{d}t + \vov \sqrt{\var_t} \;\text{d} Z_t, \label{eq:hestonb}
\end{align}
where $Z_t$ and $W_t$ are independent Brownian motions, $r$ is the riskless interest rate, $q$ is the continuous dividend yield, $\mr$ is the speed of mean reversion, $\vinf$ is the mean reversion level, $\vov$ is the volatility of the variance process, and $\rho \in [-1,1]$ represents the correlation coefficient between $S_t$ and $\var_t$. The initial conditions $S_0$ and $\var_0$ are assumed to be strictly positive. The joint process $(S, \var)$ is well known to be a time-homogeneous Markov process.

\subsection{Terminal variance, integrated variance, and asset return}
\noindent
We state several analytic properties of the variance, (conditional) integrated variance, and asset return, which are necessary for the remainder of this paper. The variance process $\var_t$ is governed by the CIR~\citep{cox1985cir} process. It is well known that the terminal variance $\var_T$ given $\var_0$ observes a noncentral chi-square distribution characterized by
\begin{equation} \label{eq:csd}
\var_T \sim \frac{e^{-\frac{\mr T}{2}}}{\phi_T(\mr)}\,\NCX\left(\delta,\var_0\,\phi_T(\mr) e^{-\frac{\mr T}{2}}\right)\qtext{for}
\delta = \frac{4\mr \vinf}{\vov^2} \qtext{and} \phi_T(\mr)=\frac{2\mr/\vov^2}{\sinh (\mr T/2)},
\end{equation}
where $\NCX(\delta,\lambda)$ denotes a noncentral chi-square random variable with degree of freedom $\delta$ and a noncentrality parameter $\lambda$. The variance process $\var_t$ cannot reach zero for $t>0$, provided that $\delta >2$ (the Feller condition). However, the Heston model parameters calibrated to the option market usually violate this condition. The mean and variance of $\var_T$ are given by
\begin{equation} \label{eq:var_mv}
E(\var_T) = \vinf + (\var_0 - \vinf)e^{-\mr T} \qtext{and}
\text{Var}(\var_T) = \frac{\vov^2}{\mr}(1-e^{-\mr T})\left[\var_0 e^{-\mr T} + \frac{\vinf}{2} (1-e^{-\mr T})\right].
\end{equation}

We define the average variance between times 0 and $T$ as 
\begin{equation} \label{eq:R}
R_{0,T} = \frac{1}{T} \int_0^T\var_t \;\text{d} t.
\end{equation}
The mean and variance of $R_{0,T}$ are found to be~\citep{ball1994stochastic}
\begin{align}
	E(R_{0,T}) &= \vinf + (\var_0 - \vinf) \frac{1-e^{- \mr T}}{\mr T}, \label{eq:R_m}\\
	\text{Var}(R_{0,T}) &= \frac{\vov^2}{\mr^2 T}\left\{ \vinf - 2(\var_0 - \vinf) e^{-\mr T} +
	\left[ \var_0 - \frac{5\vinf}2 + \left(\var_0 - \frac{\vinf}{2}\right) e^{-\mr T} \right]\frac{1-e^{-\mr T}}{\mr T}\right\}, \label{eq:R_v}
\end{align}
which provide useful insights into this model. For example, mean $E(R_{0,T})$ is the fair strike of the continuously monitored variance swap (see Section~\ref{ss:varswap}). In the uncorrelated ($\rho=0$) Heston model, $\sqrt{E(R_{0,T})}$ plays the role of the BS implied volatility in the option price approximation~\citep{ball1994stochastic}.

We also define the integrated variance between times $0$ and $T$, conditional on the initial and terminal variances, as
\begin{equation}\label{eq:iv}
 \ivt(\var_0, \var_T) = \left(\int_0^T \var_t\, \text{d}t \;\Big|\; \var_0, \var_T\right),
\end{equation}
that observes $\ivt = E(TR_{0,T}|\var_0, \var_T)$.
For notational simplicity, we simply write $\ivt$, assuming conditional dependence on $\var_0$ and $\var_T$ implicitly. Any expectation regarding $\ivt$ should be understood as a conditional expectation, $E(f(\ivt)) := E(f(\ivt)\,|\,\var_0, \var_T)$ for any function $f$, unless stated otherwise.

The exact simulation scheme is constructed, starting with the following analytic representation of the asset price process $S_t$ of the Heston model:
\begin{equation} \label{eq:ap}
  S_T = S_0 \exp \left((r-q)T - \frac{1}{2} \int_0^T \var_t \;\text{d}t + \rho \int_0^T \sqrt{\var_t} \;\text{d} Z_t + \sqrt{1-\rho^2} \int_0^T \sqrt{\var_t} \;\text{d} W_t \right).
\end{equation}
By integrating \eqref{eq:hestonb}, we obtain
\begin{equation*}
  \int_0^T \sqrt{\var_t} \;\text{d} Z_t =\frac{1}{\vov} [\var_T -\var_0 + \mr(\ivt -\vinf T)].
\end{equation*}
Conditional on $\var_T$ and $\ivt$, we deduce that the log return, $\ln(S_T/S_0)$, can be sampled from a normal distribution:
\begin{equation} \label{eq:ln}
\ln \frac{S_T}{S_0} \sim (r-q)T - \frac{\ivt}{2} + \frac{\rho}{\vov} \left[\var_T - \var_0 + \mr(\ivt - \vinf T) \right] + \Sigma_{0,T}\; Z,
\end{equation}
where $Z$ is a standard normal variate and $\Sigma_{0,T}$ is the standard deviation, as defined by
$$
\Sigma_{0,T} = \sqrt{(1-\rho^2) \ivt}.
$$
Therefore, asset price $S_T$ can be simulated as a geometric Brownian motion:
\begin{equation} \label{eq:gbm}
S_T = F_T \exp\left(\Sigma_{0,T}\, Z - \frac12 \Sigma_{0,T}^2 \right),
\end{equation}
where $F_T$ is the forward stock price, conditional on $S_0$, $\var_0$, $\var_T$, and $\ivt$:
\begin{equation} \label{eq:cond-fwd}
\begin{aligned}
F_T &= E(S_T \,|\,S_0, \var_0, \var_T, \ivt)\\
&= S_0 e^{(r-q)T} \exp \left(-\frac{\rho^2}{2} \ivt+\frac{\rho}{\vov}[\var_T -\var_0 + \mr (\ivt - \vinf T)]\right).
\end{aligned}
\end{equation}
Similar to $\ivt$, we simply write $F_T$ in the later exposition, assuming the conditional dependence on $S_0$, $\var_0$, $\var_T$, and $\ivt$ to be implicit.

Therefore, the simulation of $S_T$ given $S_0$ and $\var_0$ reduces to sampling $\var_T$ and $\ivt$ sequentially in each simulation path. As $\var_T$ can be sampled with relative ease from a noncentral chi-square distribution, the challenge in the Heston model simulation lies in sampling the conditional integrated variance $\ivt$.

\subsection{Broadie--Kaya exact simulation}
\noindent
\citet{broadie2006mcheston} perform simulation of $\ivt$ via the numerical Laplace inversion of the conditional Laplace transform of $\ivt$ via the following analytic form \citep{pitman1982decom}:
\begin{equation} \label{eq:lap-heston}
E\left( e^{-u \ivt}\right) =
\frac{\exp\left(-\frac{\var_0+\var_T}{2}\cosh(\frac{\mr_u T}{2})\phi_T(\mr_u)\right)}
{\exp\left(-\frac{\var_0+\var_T}{2}\cosh(\frac{\mr T}{2})\phi_T(\mr)\right)}
\frac{\phi_T(\mr_u)}{\phi_T(\mr)}
\frac{I_{\nu}\left( z_u \right)}{I_{\nu}\left( z \right)},
\end{equation}
where
\begin{equation} \label{eq:nu_z}
\nu = \frac{\delta}{2} - 1 = \frac{2\mr \vinf}{\vov^2} - 1,\;\;
z = \sqrt{\var_0\,\var_T} \phi_T(\mr),\;\; z_u = \sqrt{\var_0\,\var_T} \phi_T(\mr_u),
\;\;
\mr_u = \sqrt{\mr^2 + 2\vov^2 u},
\end{equation}
and $I_\nu(z)$ is a modified Bessel function of the first kind:
\begin{equation} \label{eq:I_v}
	I_\nu(z) = \sum_{k=0}^{\infty} \frac{(z/2)^{\nu + 2k}}{k! \,\Gamma(k+\nu+1)}.
\end{equation}
This sampling procedure for $\ivt$ is very time-consuming since the Laplace inversion algorithm requires cumbersome numerical evaluations of $I_\nu(z)$ over the grid of $z$ values. Moreover, such evaluation must be performed for each simulation path since the conditional Laplace transform depends on $\var_T$.

\subsection{Gamma expansion derived from the Bessel bridge decomposition} \label{ss:ge}
\noindent
A significant improvement in computational efficiency can be achieved if the numerical inversion of the Laplace transform can be avoided. Based on the Bessel bridge decomposition~\citep{pitman1982decom}, \citet{glasserman2011gamma} manage to express
the conditional integrated variance $\ivt$ as an infinite sum of the gamma random variables.

Let $\POIS(\lambda)$ denote the Poisson random variable with rate $\lambda$, and $\GAM(\alpha)$ denote the standard (i.e., unit scale) gamma random variable with shape parameter $\alpha$. The probability mass function of $\POIS(\lambda)$ and probability density function of $\GAM(\alpha)$ are respectively given by
$$ P_\POIS(j;\lambda) = \frac{\lambda^j\, e^{-\lambda}}{j!} \;\; (j=0,1,\ldots) \qtext{and}
 f_{\GAM}(x;\alpha) = \frac{x^{\alpha-1}e^{-x}}{\Gamma(\alpha)} \;\; (x>0).
$$
Then, $\ivt$ can be expressed as~\citep[Theorem~2.2]{glasserman2011gamma}
\begin{equation} \label{eq:is}
\ivt \sim X + Z_{\delta/2} + \sum_{j=1}^{\eta_{0,T}} Z_2^{(j)},
\end{equation}
where
\begin{gather*}
	X \sim \sum_{k=1}^{\infty} \frac{1}{\gamma_k} \GAM(n_k) \qtext{and} Z_\alpha \sim \sum_{k=1}^{\infty} \frac{1}{\gamma_k} \GAM(\alpha) \qtext{for} n_k(\var_0,\var_T) \sim \POIS( (\var_0+\var_T)\lambda_k )\; \text{i.i.d.},\\
	\lambda_k = \frac{16k^2 \pi^2}{\vov^2 T (\mr^2 T^2 + 4k^2 \pi^2)} \qtext{and}
\gamma_k = \frac{\mr^2 T^2 + 4k^2 \pi^2}{2\vov^2 T^2},
\end{gather*}
and $Z_\alpha^{(j)}$ denotes independent identically distributed copies of $Z_\alpha$. The last term in the infinite series \eqref{eq:is} involves the summation of terms whose number equals to a random variable $\eta_{0,T}$ with dependence on $\var_0$ and $\var_T$, where
\begin{equation} \label{eq:bes}
	\eta_{0,T}(\var_0,\var_T) \sim \BES(\nu, z),
\end{equation}
where $\nu$ and $z$ are defined in \eqref{eq:nu_z}. The Bessel random variable, $\eta \sim \BES(\nu, z)$ with $\nu >-1$ and $z>0$, takes non-negative integer values. Its probability mass function is given by the normalized coefficients of $I_\nu(z)$ in \eqref{eq:I_v}:
\begin{equation} \label{eq:bes_p}
P_\BES(j;\nu,z) = \frac{(z/2)^{2j+\nu}}{I_\nu(z)\,j! \,\Gamma(j+\nu+1)} \quad (j=0,1,\ldots).
\end{equation}

In the actual numerical implementation of evaluating the infinite gamma series, it is necessary to evaluate the sum of only finite $K$ terms and properly approximate the truncated terms, 
$$ X^K \sim \sum_{k=K+1}^{\infty} \frac{1}{\gamma_k} \GAM(n_k) \qtext{and} Z_\alpha^K \sim \sum_{k=K+1}^{\infty} \frac{1}{\gamma_k} \GAM(\alpha).
$$
\citet[Proposition 3.2]{glasserman2011gamma} propose to approximate the three truncated terms of $X$, $Z_{\delta/2}$, and $Z_2$ with gamma random variables via matching mean and variance. The mean and variance of the truncated terms are available analytically. \citet{glasserman2011gamma} obtain the mean and variance of $X$ and $Z_\alpha$ as follow:
\begin{equation} \label{eq:XZ-mv}
\begin{aligned}
	E(X)& = (\var_0 + \var_T)\sum_{k=1}^{\infty} \frac{\lambda_k}{\gamma_k} = (\var_0 + \var_T)\sum_{k=1}^{\infty} \frac{32 \pi^2 k^2 \, T}{(\mr^2 T^2 + 4k^2 \pi^2)^2} = (\var_0 + \var_T) m_X T, \\
	\text{Var}(X) &= (\var_0 + \var_T)\sum_{k=1}^{\infty} \frac{2\lambda_k}{\gamma_k^2} = (\var_0 + \var_T)\sum_{k=1}^{\infty} \frac{128 \pi^2 k^2 \, \vov^2 T^3}{(\mr^2 T^2 + 4k^2 \pi^2)^3} = (\var_0 + \var_T) v_X \vov^2 T^3, \\
	E(Z_\alpha) &= \sum_{k=1}^{\infty} \frac{\alpha}{\gamma_k} = \sum_{k=1}^{\infty} \frac{2\alpha \vov^2 T^2}{\mr^2 T^2 + 4k^2 \pi^2} = \alpha m_Z \vov^2 T^2, \\
	\text{Var}(Z_\alpha) &= \sum_{k=1}^{\infty} \frac{\alpha}{\gamma_k^2} = \sum_{k=1}^{\infty} \frac{4 \alpha \vov^4 T^4}{(\mr^2 T^2 + 4k^2 \pi^2)^2} = \alpha v_Z \vov^4 T^4,
\end{aligned}
\end{equation}
where
\begin{equation} \label{eq:mv}
\begin{gathered}
	m_X = \frac{c_1 - a c_2}{2a}, \quad v_X = \frac{c_1 + a c_2 - 2a^2 c_1c_2}{8a^3}, \quad
	m_Z = \frac{a c_1 - 1}{4a^2}, \quad v_Z = \frac{a c_1 + a^2 c_2 - 2}{16a^4}\\
	a = \frac{\mr T}{2},\quad c_1 = \frac1{\tanh a}, \qtext{and} c_2 = \frac1{\sinh^2 a}.
\end{gathered}
\end{equation}
The mean and variance of $X^K$ and $Z_\alpha^K$ are then obtained by subtracting the first $K$ terms from those of $X$ and $Z_\alpha$, respectively. We obtain 
$$ E(X^K) = E(X) - (\var_0 + \var_T)\sum_{k=1}^K \frac{\lambda_k}{\gamma_k} \qtext{and} E(Z_\alpha^K) = E(Z_\alpha) - \sum_{k=1}^K \frac{\alpha}{\gamma_k}. $$

It is obvious that $K$ serves as a parameter controlling the accuracy of the simulation scheme; a higher $K$ implies a smaller error at the tradeoff of a higher computational cost. Note that the numerical implementation of the GE scheme \eqref{eq:is} eventually requires $(2+\eta_{0,T})(K+1)$ gamma random variables per simulation path. 

The GE representation of $\ivt$ proposed by \citet{glasserman2011gamma} avoids the tedious numerical inversion of the conditional Laplace transform of $\ivt$ as implemented by \citet{broadie2006mcheston}. However, substantial computational effort is still required to simulate the Bessel random variable $\eta_{0,T}$, as it involves $I_\nu(z)$ as well.\footnote{See \citet[Table 4]{glasserman2011gamma} for the computation time required for sampling the Bessel random variables.}

\subsection{Inverse Gaussian approximation based on matching moments} \label{ss:ig}
\noindent
\citet{tse2013lowbias} propose a low-bias simulation scheme, where $\ivt$ is approximated by an IG variate with matching mean and variance of $\ivt$. They argue that the IG variable is the best candidate for approximating $\ivt$ because the two converge in distribution as $T\to \infty$.
Let $\IG(\mu,\lambda)$ denote the IG random variable with the parameters $\mu$ and $\lambda$. The density function of $\IG(\mu,\lambda)$ takes the form
\begin{equation} \label{eq:IGdis}
f_{\IG}(x;\mu,\lambda)= \sqrt{\frac{\lambda}{2 \pi x^3}} \exp \left( -\frac{\lambda (x-\mu)^2}{2\mu^2 x} \right) \qtext{for} \mu>0,\; \lambda >0, \qtext[\;]{and} x>0,
\end{equation}
and the mean and variance of $\IG(\mu, \lambda)$ are given by $\mu$ and $\mu^3/\lambda$, respectively. Parameters $\mu$ and $\lambda$ are determined by matching the mean and variance of $\ivt$:
$$ \mu = E(\ivt) \qtext{and} \lambda = \frac{E(\ivt)^3}{\text{Var}(\ivt)}.
$$
The exact mean and variance of $\ivt$ are found to be~\citep[Proposition~3.1]{tse2013lowbias}:
\begin{equation} \label{eq:iv-mv}
	\begin{aligned}
		E(\ivt) &= E(X) + E(Z_{\delta/2}) + E(\eta_{0,T})E(Z_2)\\
		\text{Var}(\ivt) &= \text{Var}(X) + \text{Var}(Z_{\delta/2}) + E(\eta_{0,T})\text{Var}(Z_2) + \text{Var}(\eta_{0,T}) E(Z_2)^2,
	\end{aligned}
\end{equation}
where the mean and variance of $X$ and $Z_\alpha$ are given in \eqref{eq:XZ-mv} and 
\begin{equation} \label{eq:eta-mv}
E(\eta_{0,T}) = \frac{z\,I_{\nu + 1}(z)}{2I_{\nu}(z)} \qtext{and}
\text{Var}(\eta_{0,T}) = \frac{z^2\,I_{\nu + 2}(z)}{4 I_\nu(z)} + E(\eta_{0,T}) - E(\eta_{0,T})^2.
\end{equation}
Once $\mu$ and $\lambda$ have been determined, it is trivial to sample the $\IG(\mu,\lambda)$ variate from the algorithm of \citet{michael1976gen}.

Note that the error-controlling parameter, similar to $K$ in the GE scheme, is not present in the IG approximation. As \citet{tse2013lowbias} demonstrate, the only way to reduce the simulation bias is to decrease the time interval by increasing the number of simulation steps. Therefore, we can classify their method as a low-bias scheme rather than an exact simulation scheme.

However, to obtain $E(\ivt)$ and $\text{Var}(\ivt)$, the modified Bessel functions (i.e., $I_{\nu}(z)$, $I_{\nu + 1}(z)$, and $I_{\nu + 2}(z)$) must be evaluated for each path, which causes some computational burden. To avoid this problem, \citet{tse2013lowbias} pre-compute $E(\eta_{0,T})$ and $\text{Var}(\eta_{0,T})$ for a grid of equally spaced $\var_0 \var_T$ values, thanks to the property that $\eta_{0,T}$ depends on $\var_0$ and $\var_T$ via $z=\sqrt{\var_0 \var_T}\, \phi_T(\mr)$. Interpolation within the tabulated values is then performed. In addition, they propose the interpolation--Poisson--zero (IPZ) scheme for faster sampling of the noncentral chi-square variate for $\var_T$ (see more comments in Section~\ref{s:num}).

\subsection{Time discretization with QE approximation, trapezoidal rule, and martingale correction} \label{ss:qe}
\noindent
For pricing path-dependent derivatives such as the Asian options and variance swaps, it is necessary to sample asset prices at frequent time points. For this purpose, the simulation schemes based on time discretization (e.g., Euler or Milstein) become competitive in terms of the accuracy-speed tradeoff over the exact (e.g., GE) or low-bias (e.g., IG) schemes. The time discretization scheme is distinguished from those discussed earlier in the sense that the conditional variance is approximated as a deterministic value rather than as a random variate from the distribution of $\ivt$, be it exact or approximate. Therefore, such procedure must be applied to small time intervals successively. Despite this limitation, the computational cost per time step is lower than that of other schemes. Therefore, time discretization is the preferred scheme if such a short time step is required for the derivatives with high monitoring frequency.

The Euler and Milstein schemes are direct time discretizations of the stochastic differential equation. In the Heston model, however, these two simple schemes are notorious for failing with a large bias owing to the square root process for variance. Among the numerous studies on time discretization schemes under the Heston model \citep{kahl2006fast,lord2010comparison}, the QE scheme proposed by \citet{andersen2008simple} has been widely recognized as the best scheme~\citep{vanhaastrecht2010efficient}. The QE scheme is briefly reviewed below:

\vspace{1ex}
\noindent \textbf{Sampling variance: QE approximation}. Under time discretization, the life span of the derivative, $[0,T]$, is divided into $N$ equal time intervals of size $\dt$ ($T = N\dt$). The monitoring points within the time interval are specified as $t_i=i\dt$  ($i=0,1,\cdots,N$). 
As it is more concise to use index $i$ than time $t_i$, we change the notation convention in this section (and in Section~\ref{ss:pois-dt} later) by using $i$ in subscripts. For example, we use
$$ \var_i := \var_{t_i} \qtext{and} \ivstep := I_{t_i,t_{i+1}} \;\; (\ivi := \ivt).
$$
In the QE scheme, $\var_{i+1}$ given $\var_i$ is sampled using approximate functional forms. First, using \eqref{eq:var_mv}, we calculate the ratio, $\psi = \text{Var}(\var_{i+1}|\var_i)/E(\var_{i+1}|\var_i)^2$, as a proxy for the probability that $\var_{i+1}$ hits the origin. Then, the simulation is split into two cases depending on $\psi$: 
\begin{equation} \label{eq:qe}
\var_{i+1} = 
\begin{cases} 
a (b+Z)^2 & \qtext{if} \psi \le 1.5 \\
\frac{1}{\beta} \ln\left(\frac{1-p}{1-U}\right) \boldsymbol{1}_{U>p} & \qtext{if} \psi > 1.5
\end{cases},
\end{equation}
where $Z$ and $U$ are standard normal and uniform random variables, respectively; $\boldsymbol{1}_x$ is the indicator function; and the coefficients $a$, $b$, $\beta$, and $p$ are determined to match $E(\var_{i+1}\,|\,\var_i)$ and $\text{Var}(\var_{i+1}\,|\,\var_i)$ in each case.

\vspace{1ex}
\noindent \textbf{Trapezoidal rule}. After simulating $\var_{i+1}$ from $\var_i$, \citet{andersen2008simple} adopts a simple trapezoidal rule to approximate the conditional integrated variance $\ivstep$ as follows:
\begin{equation} \label{eq:tz}
\ivstep^\text{TZ} = (\var_i + \var_{i+1})\frac{\dt}{2}.
\end{equation}
Given the simulation path $\{\var_i: i=0,1,\cdots,N\}$, $\ivi$ over the entire period is approximated by
\begin{equation} \label{eq:tz-sum}
\ivi^\text{TZ} = \sum_{i=0}^{N-1}\ivstep^\text{TZ} = (\var_0 + 2\var_1 + \cdots + 2\var_{N-1} + \var_N)\frac{\dt}{2}.
\end{equation}
As previously noted, $\ivstep^\text{TZ}$ (conditional on $\var_i$ and $\var_{i+1}$) is a deterministic value. This feature signifies an important difference with compared with the GE and IG schemes, where $\ivstep$ is sampled as a random variable. 

\vspace{1ex}
\noindent \textbf{Martingale correction}. Finally, to ensure the martingale condition, $S_i = e^{(q-r)h} E( F_{i+1}|S_i)$, \citet{andersen2008simple} modifies the conditional forward price in \eqref{eq:cond-fwd} by adding a correction term, $M_{i,i+1}^\text{QE}$:
$$
F_{i+1} = S_i e^{(r-q)h} \exp \left(-\frac{\rho^2}{2} \ivstep^\text{TZ} + \frac{\rho}{\vov}[\var_{i+1} -\var_i + \mr (\ivstep^\text{TZ} - \vinf h)] + M_{i,i+1}^\text{QE}\right).
$$
The martingale correction term, $M_{i,i+1}^\text{QE}$, is analytically determined from the approximation function~\eqref{eq:qe}:
$$
M_{i,i+1}^\text{QE} = \frac{\rho\mr\theta}{\vov}h - A_2 \var_i + 
\begin{cases} 
-\frac{A_1 b^2 a}{1-2A_1 a} + \frac12\ln(1-2A_1 a) & \text{if}\;\; \psi \le 1.5 \\
-\ln\left(p + \frac{\beta(1-p)}{\beta-A_1}\right) & \text{if}\;\; \psi > 1.5
\end{cases} \qtext{and}
A_{1,2} = \frac{\rho\dt}{4} \left(\frac{2\mr}{\vov} - \rho \right) \pm \frac{\rho}{\vov},
$$
where $A_1$ and $A_2$ takes $+$ and $-$ respectively, and $a$, $b$, $\beta$, and $p$ are the coefficients used in the QE step in \eqref{eq:qe}.

\section{Poisson conditioning and enhanced simulation schemes} \label{s:pois}
\noindent
In this section, we construct efficient simulation schemes for the Heston model based on the key observations of Poisson conditioning, which simplifies the model formulation. The primary merit of Poisson conditioning is that it removes the use of Bessel functions and Bessel random variables, the numerical evaluation of which is computationally demanding.

\subsection{Poisson conditioning}
\noindent
It is well known that the noncentral chi-square random variable $\NCX(\delta, \gamma)$ is a Poisson mixture of an ordinary ($\gamma=0$) chi-square variables. Recall that an ordinary chi-square distribution is a special case of a gamma distribution, $\NCX(\delta, 0)\sim2\GAM(\delta/2)$: 
$$
	\NCX(\delta, \gamma) \sim \NCX\left(\delta + 2 \POIS\left(\frac{\gamma}{2}\right),0\right) \sim 2\,\GAM\left(\frac{\delta}{2} + \POIS\left(\frac{\gamma}{2}\right)\right).
$$
With these properties, the exact simulation scheme of $\var_T$ in \eqref{eq:csd} can be alternatively recast as the following Poisson mixture gamma scheme:
\begin{equation} \label{eq:var_T}
\mu_0 \sim \POIS\left(\frac{\var_0\,\phi_T(\mr) e^{-\frac{\mr T}{2}}}{2}\right), \qtext{so that} \var_T \sim \frac{2e^{-\frac{\mr T}{2}}}{\phi_T(\mr)}\GAM\left(\frac{\delta}{2} + \mu_0\right).
\end{equation}
Note that $\var_T$ depends on $\var_0$ via $\mu_0$. This property is well known and often used in the Heston simulation literature~\citep[see][]{vanhaastrecht2010efficient,glasserman2011gamma,tse2013lowbias}. Our new simulation schemes also use this Poisson mixture property to simulate $\var_T$. Unlike other schemes, the intermediate Poisson variable $\mu_0$ is also essential for simulating $\ivt$, in addition to $\var_T$ in our proposed scheme, as detailed below.

The choice of \eqref{eq:var_T} for simulating $\var_T$ does not significantly increase the computation time. Our numerical tests with public numerical library show that \eqref{eq:var_T} is marginally slower than simulating the noncentral chi-square variable directly. The small difference is attributed to the generation of the Poisson variable $\mu_0$. Interestingly, we also observe that \eqref{eq:var_T} is comparable to Andersen's QE procedure for generating $\var_T$ (see more comments in Section~\ref{ss:pois-dt}).

Our key observation for Poisson conditioning is the link between the Poisson random variable $\mu_0$ in \eqref{eq:var_T} and the Bessel random variable $\eta_{0,T}$ in \eqref{eq:bes}. According to \citet[Eq.~(5.j)]{pitman1982decom}, the Bessel random variable $\eta \sim \BES(\nu, z)$ can be represented alternatively as a conditional Poisson random variable:
$$ \mu\sim \POIS(\lambda) \qtext{conditional on} \GAM(\nu + 1 + \mu) = \frac{z^2}{4\lambda},
$$
where $\lambda$ is a positive rate parameter that can be chosen arbitrarily. This can be derived from the observation that the joint probability of $\mu$ and $\GAM(\nu + 1 + \mu)$ is the product of the two probabilities\footnote{This is not because $\mu$ and $\GAM(\nu + 1 + \mu)$ are independent, but rather $\mu$ is the parameter of $\GAM(\nu + 1 + \mu)$. Given $\mu=j$, $\GAM(\nu + 1 + \mu)$ is a standard Gamma variate. We use the conditional probability formula:
\begin{align*}
\text{Prob}&\left(\mu=k \;\cap\; \GAM(\nu + 1 + \mu) = y\right) \\
&= \text{Prob}(\mu=k)\; \text{Prob}\left(\GAM(\nu + 1 + \mu) = y \;|\; \mu=k \right)\\
&= \text{Prob}(\mu=k)\; \text{Prob}\left(\GAM(\nu + 1 + k) = y \right)\\
&= P_\POIS(k;\lambda)\; f_{\GAM}\!\left(y; \nu + 1 + k \right),
\end{align*} 
where $y=\frac{z^2}{4\lambda}$.}, and it is proportional to the probability mass function of $\eta$ in \eqref{eq:bes_p}:
$$
P_\POIS(k;\lambda)\; f_{\GAM}\!\left(\frac{z^2}{4\lambda}; \nu + 1 + k \right) = 
\frac{\lambda^k\, e^{-\lambda}}{k!} \frac{\left(\frac{z^2}{4\lambda}\right)^{\nu + k}e^{-\frac{z^2}{4\lambda}}}{\Gamma(\nu + 1 + k)} =  \left(\frac{z}{2\lambda}\right)^{\nu}\!e^{-\lambda-\frac{z^2}{4\lambda}}\,
\frac{(z/2)^{2k+\nu}}{k! \,\Gamma(k+\nu+1)}.
$$
Therefore, the alternative representation is obtained as follows:
$$ \text{Prob}\left(\mu=j \,\Big|\, \GAM(\nu + 1 + \mu) = \frac{z^2}{4\lambda}\right) = \frac{P_\POIS(j;\lambda)\; f_{\GAM}\!\left(\frac{z^2}{4\lambda}; \nu + 1 + j \right)}{\sum_{k=0}^\infty P_\POIS(k;\lambda)\; f_{\GAM}\!\left(\frac{z^2}{4\lambda}; \nu + 1 + k \right)} = P_\BES(j;\nu,z).
$$

Instead of taking $\lambda=1$ as in \citet[Remark 2.3]{glasserman2011gamma}, we can achieve a useful representation that resembles that of $\mu_0$ by judiciously choosing $\lambda = \frac{\var_0\,\phi_T(\mr) }{2}e^{-\frac{\mr T}{2}}$ in the context of the Heston model. By substituting $\nu=\frac{\delta}{2}-1$ and $z=\sqrt{\var_0 \var_T}\, \phi_T(\mr)$, $\eta_{0,T}\sim \BES(\nu, z)$ is equivalent to the conditional Poisson variable:
$$ \mu \sim \POIS\left(\frac{\var_0\,\phi_T(\mr) e^{-\frac{\mr T}{2}}}{2}\right) \qtext{conditional on} \GAM\left(\frac{\delta}{2} + \mu\right) = \frac{z^2}{4 \lambda} = \frac{\var_T \phi_T(\mr)}{2e^{-\frac{\mr T}{2}} }.
$$
Consequently, $\mu$ coincides with $\mu_0$ in \eqref{eq:var_T}. Simultaneously, we observe that the required condition for $\mu$ is equivalent to the formula for $\var_T$ in \eqref{eq:var_T}:
$$\var_T = \frac{2e^{-\frac{\mr T}{2}}}{\phi_T(\mr)}\GAM\left(\frac{\delta}{2}+\mu\right).$$
Therefore, we conclude that $\eta_{0,T}$ is equivalent to $\mu_0$ conditional on the terminal variance $\var_T$:
\begin{equation} \label{eq:eta}
\eta_{0,T} \;\;\sim\;\; \mu_0 \;\Big|\; \var_T = \frac{2e^{-\frac{\mr T}{2}}}{\phi_T(\mr)}\GAM\left(\frac{\delta}{2} + \mu_0\right).
\end{equation}

This implies that the joint distribution of $(\var_T, \eta_{0,T})$ is equivalent to that of $(\var_T, \mu_0)$ as long as $\mu_0$ and $\var_T$ follow the relation in \eqref{eq:var_T}. Therefore, $\eta_{0,T}$ can be simply replaced by $\mu_0$ when sampling $\ivt$. As a result, the cumbersome evaluation of the modified Bessel function is completely avoided under our Poisson conditioning framework. Under Poisson conditioning, \eqref{eq:is} of the GE scheme can be further simplified to
\begin{equation} \label{eq:is2}
	\ivt\,|\,\mu_0 \sim X + Z_{\delta/2} + \sum_{j=1}^{\mu_0} Z_2^{(j)}.
\end{equation}
This is the key result of an efficient approach coined as Poisson conditioning. In the remainder of this section, we show how various Heston simulation schemes can be simplified under the Poisson conditioning framework.

\subsection{Poisson-conditioned IG approximation} \label{ss:pois-ig}
\noindent
First, we enhance \citet{tse2013lowbias}'s IG approximation using Poisson conditioning. The Poisson-conditioned IG approximation in this section is a special case of the more general Poisson-conditioned GE scheme introduced in the next section. Nevertheless, we first discuss this scheme since the mean and variance of $\ivt\,|\,\mu_0$ to be derived in \eqref{eq:iv-pois-mv} will be used later.

Recall that the modified Bessel function in \eqref{eq:eta-mv} is the computational bottleneck in the original IG scheme. Under Poisson conditioning, we observe $E(\eta_{0,T}| \mu_0)=\mu_0$ and $\text{Var}(\eta_{0,T} | \mu_0)=0$ because $\eta_{0,T}$ is now conditioned as $\mu_0$. Consequently, the mean and variance of $\ivt\,|\,\mu_0$ can be simplified as follows (see \eqref{eq:iv-mv} for comparison):
\begin{equation} \label{eq:iv-pois-mv}
\begin{aligned}
	E(\ivt\,|\,\mu_0) &= E(X) + E(Z_{\delta/2}) + \mu_0 E(Z_2)
	= (\var_0 + \var_T) m_X T + \left(\frac{\delta}{2} + 2\mu_0\right) m_Z \vov^2 T^2, \\
	\text{Var}(\ivt\,|\,\mu_0) &= \text{Var}(X) + \text{Var}(Z_{\delta/2}) + \mu_0\text{Var}(Z_2)
	= (\var_0 + \var_T) v_X \vov^2 T^3 + \left(\frac{\delta}{2} + 2\mu_0\right) v_Z \vov^4 T^4,
\end{aligned}
\end{equation}
which are expressed in terms of elementary functions, with no reference to the Bessel function anymore. Therefore, we expect significant speedup in calculating the conditional mean and variance of $\ivt$ for moment matching with the IG variate.

\subsection{Poisson-conditioned GE scheme} \label{ss:pois-ge}
\noindent
We present the main results of the Poisson-conditioned GE scheme. For the simulation of $\ivt\,|\,\mu_0$ in \eqref{eq:is2}, we apply two additional enhancements.

\vspace{1ex}
\noindent
\textbf{Aggregating gamma random variables}. 
First, using the additive property of the independent gamma variables, we obtain
\begin{equation*}
	\GAM(\alpha_1) + \GAM(\alpha_2) \sim \GAM(\alpha_1+\alpha_2),
\end{equation*}
so that we can combine the $(2+\mu_0)$ gamma variables in \eqref{eq:is2} as follows:
\begin{equation} \label{eq:add}
	\ivt\,|\,\mu_0 \sim \sum_{k=1}^{\infty} \left[\frac{1}{\gamma_k} \GAM(n_k) + \frac{1}{\gamma_k} \GAM(\delta/2) + \sum_{j=1}^{\mu_0} \frac{1}{\gamma_k} \GAM(2) \right]
	\sim \sum_{k=1}^{\infty} \frac{1}{\gamma_k} \GAM\left(n_k + \frac{\delta}{2} + 2\mu_0 \right).
\end{equation}
The $(2+\mu_0)$ infinite series is now converted into a single series. This reduces the number of required gamma variates by more than one-third of the original GE scheme.

\vspace{1ex}
\noindent
\textbf{Series truncation with the IG approximation}. 
In the second enhancement, we improve the accuracy of truncating the infinite series \eqref{eq:add} using the IG approximation instead of the gamma approximation in \citet{glasserman2011gamma}. 
Owing to the aggregated gamma variable, we apply the truncation to the whole $\ivt$ term rather than each of $X$, $Z_{\delta/2}$, or $Z_2$ terms, as in \citet{glasserman2011gamma}. As \citet{tse2013lowbias} have already shown the effectiveness of the IG variate for approximating the entire $\ivt$, it is natural to expect that the IG approximation also works for the remaining terms of $\ivt$. Our numerical tests confirm this result. 

Let $\ivt^{K}\,|\,\mu_0$ denote the remainder of the first $K$ terms of \eqref{eq:add} under Poisson conditioning.
\begin{equation} \label{eq:RK}
	\ivt^{K}\,|\,\mu_0 \sim \sum_{k=K+1}^{\infty} \frac{1}{\gamma_k} \GAM\left(n_k + \frac{\delta}{2} + 2\mu_0\right).
\end{equation}
We approximate $\ivt^{K}\,|\,\mu_0$ using an IG random variable via matching mean and variance. The mean and variance of $\ivt^{K}\,|\,\mu_0$ can be easily obtained by subtracting those of the first $K$ terms from \eqref{eq:iv-pois-mv}:
\begin{equation} \label{eq:iv-truc-mv}
\begin{aligned}
	E(\ivt^{K}\,|\,\mu_0) &= (\var_0 + \var_T) \left(m_X T - \sum_{k=1}^K \frac{\lambda_k}{\gamma_k} \right)+ \left(\frac{\delta}{2} + 2\mu_0\right) \left(m_Z \vov^2 T^2 - \sum_{k=1}^K \frac1{\gamma_k} \right),\\
	\text{Var}(\ivt^{K}\,|\,\mu_0) &= (\var_0 + \var_T) \left(v_X \vov^2 T^3 - \sum_{k=1}^K \frac{2\lambda_k}{\gamma_k^2} \right) + \left(\frac{\delta}{2} + 2\mu_0\right) \left(v_Z \vov^4 T^4 - \sum_{k=1}^K \frac1{\gamma_k^2}\right).
\end{aligned}
\end{equation}
As in \eqref{eq:iv-pois-mv}, there is no reference to the Bessel functions.

Note that the case where $K=0$ in our new scheme is reduced to the Poisson-conditioned IG scheme in Section~\ref{ss:pois-ig}. However, this is not the case with the original GE and IG schemes, in terms of both the type and number of random variables. The $K=0$ case of the GE scheme consists of ($2+\eta_{0,T}$) IG variates approximating $X$, $Z_{\delta/2}$, and copies of $Z_2$ whereas the IG scheme uses one IG variate.
Therefore, the proposed Poisson-conditioned GE scheme can be used with flexibility as either an exact scheme ($K\ge 1$ with one time step) or a low-bias scheme ($K=0$ with multiple time steps).

Finally, the simulation of $\var_T$ and $\ivt$ under Poisson conditioning can be succinctly performed using the following simplified steps:

\vspace{1ex}
\noindent
\textbf{Simulation steps of the Poisson-conditioned GE scheme}. 
\begin{description}
	\item[Step 1] Given $\var_0$, draw $\mu_0$ as
	\begin{equation*}
		\mu_0 \sim \POIS\left(\frac{\var_0\,\phi_T(\mr) e^{-\frac{\mr T}{2}}}{2}\right). 
	\end{equation*}
	\item[Step 2] Given $\mu_0$, draw $\var_T$ as
	\begin{equation*}
		\var_T \sim \frac{2e^{-\frac{\mr T}{2}}}{\phi_T(\mr)}\GAM\left(\frac{\delta}{2}+\mu_0\right).
	\end{equation*}
	\item[Step 3] Given $\var_0$, $\var_T$, and $\mu_0$, draw $\ivt$ as
	\begin{equation} \label{eq:I}
		\ivt \sim \sum_{k=1}^{K} \frac{1}{\gamma_k} \GAM \left(n_k + \frac{\delta}{2} + 2\mu_0 \right) + \IG(\lambda_K, \mu_K),
	\end{equation}
	where $n_k \sim \POIS( (\var_0+\var_T)\lambda_k)$ and $\IG(\lambda_K, \mu_K)$ is an IG variate with the mean and variance matched to \eqref{eq:iv-truc-mv}.
	\item[Step 4] Given $\var_0$, $\var_T$, $\ivt$, and $S_0$, draw $S_T$ from \eqref{eq:ln}.
\end{description}
Here, both $\mu_0$ and $n_k$ are sampled from the Poisson distribution, which is computationally trivial. The simulation of $\var_T$ and $\ivt$ are indifferent to whether the Feller condition for the CIR variance process is satisfied.

\subsection{Poisson-conditioned Broadie--Kaya Laplace inversion scheme}
\noindent
Although the \citet{broadie2006mcheston} procedure of numerical inversion of the conditional Laplace transform of $\ivt$ is not recommended as an efficient numerical procedure, it may be instructive to show that Poisson conditioning can also simplify their numerical procedure. 

From the Poisson-conditioning decomposition \eqref{eq:is2}, the Laplace transform of $\ivt$ conditional on $\mu_0$ is obtained as follows:
\begin{equation} \label{eq:newlap}
\begin{aligned}
	E\left( e^{-u \ivt} \,\Big|\, \mu_0 \right) &= E\left( e^{-u X} \right) \, E\left( e^{-u Z_{\delta/2}} \right) \, E\left( e^{-u Z_2} \right)^{\mu_0}\\
	&= \frac{\exp\left(-\frac{\var_0+\var_T}{2}\cosh(\frac{\mr_u T}{2})\phi_T(\mr_u)\right)}
{\exp\left(-\frac{\var_0+\var_T}{2}\cosh(\frac{\mr T}{2})\phi_T(\mr)\right)}
\left[\frac{\phi_T(\mr_u)}{\phi_T(\mr)}\right]^{\delta/2 + 2\mu_0},
\end{aligned}
\end{equation}
where $\mr_u$ is defined in \eqref{eq:nu_z}. This is a direct consequence of \citet[Lemma 2.4]{glasserman2011gamma}, which observes
$$ E\left( e^{-u X} \right) = \frac{\exp\left(-\frac{\var_0+\var_T}{2}\cosh(\frac{\mr_u T}{2})\phi_T(\mr_u)\right)}
{\exp\left(-\frac{\var_0+\var_T}{2}\cosh(\frac{\mr T}{2})\phi_T(\mr)\right)} 
\qtext{and} E\left( e^{-u Z_\alpha} \right) = \left[\frac{\phi_T(\mr_u)}{\phi_T(\mr)}\right]^\alpha. $$
We expect to gain a significant computational benefit when compared with the original Broadie--Kaya algorithm since it no longer requires numerical evaluation of the modified Bessel functions, as in \eqref{eq:lap-heston}.

If we use the Laplace inversion of \eqref{eq:newlap} instead of the Poisson-conditioned GE scheme in \eqref{eq:I}, Step 3 can be replaced by the following alternative procedure:
\begin{description}
	\item[Step 3'] Given $\var_0$, $\var_T$, and $\mu_0$, draw $\ivt$ from the cumulative distribution function obtained from the Laplace inversion of \eqref{eq:newlap}.
\end{description}
However, this alternative Broadie--Kaya scheme is still slower than the Poisson-conditioned GE scheme. We do not include this alternative scheme in our numerical test.

\vspace{1ex}
\noindent
\textit{Remark}

\noindent
The original Broadie--Kaya conditional Laplace transform \eqref{eq:lap-heston} can also be derived in a similar manner from the original decomposition of $\ivt$ in \eqref{eq:is}. This serves as an alternative proof of \eqref{eq:is}, without resorting to the Bessel bridge decomposition used in \citet{glasserman2011gamma}. We outline the derivation below, since it also explains why \eqref{eq:lap-heston} contains $I_\nu(z)$ and how $\eta_{0,T}$ is related to $I_\nu(z)$.

The Laplace transform of \eqref{eq:is} is expressed as
$$ E\left( e^{-u \ivt} \right) = E\left( e^{-u X} \right) \, E\left( e^{-u Z_{\delta/2}} \right) \, E\left( E\left( e^{-u Z_2} \right)^{\eta_{0,T}} \right).
$$
It is already known that the Laplace transform of $X$ corresponds to the first term in \eqref{eq:lap-heston}. For the second and third terms, we use the probability-weighted average over $\eta_{0,T}$.
Based on the probability mass function of $\eta_{0,T}$ in \eqref{eq:bes_p} and series expansion of $I_\nu(z)$ in \eqref{eq:I_v}, we deduce the remaining terms in \eqref{eq:lap-heston}. Finally, we obtain
\begin{equation} \label{eq:tech}
\begin{aligned}
	E\left( e^{-u Z_{\delta/2}} \right) \, E\left( E\left( e^{-u Z_2} \right)^{\eta_{0,T}} \right) =&  \left[\frac{\phi_T(\mr_u)}{\phi_T(\mr)}\right]^{\delta/2} \sum_{j=0}^{\infty} P(\eta_{0,T}=j) \left[\frac{\phi_T(\mr_u)}{\phi_T(\mr)}\right]^{2j} \\
	=& \frac{\phi_T(\mr_u)}{\phi_T(\mr)} \sum_{j=0}^{\infty} \frac{[\sqrt{\var_0 \var_T}\, \phi_T(\mr)/2]^{\delta/2 -1 + 2j}}{I_\nu(\sqrt{\var_0 \var_T}\, \phi_T(\mr)) j! \,\Gamma(j+\nu+1)} \left[\frac{\phi_T(\mr_u)}{\phi_T(\mr)}\right]^{\delta/2 -1 + 2j} \\
	=& \frac{\phi_T(\mr_u)}{\phi_T(\mr)} \sum_{j=0}^{\infty} \frac{[z_u/2]^{\nu + 2j}}{I_\nu(z) j! \,\Gamma(j+\nu+1)}
	= \frac{\phi_T(\mr_u)}{\phi_T(\mr)} \frac{I_\nu(z_u)}{I_\nu(z)},
\end{aligned}
\end{equation}
where $\mr_u$ and $z_u$ are defined in \eqref{eq:nu_z}.

\subsection{Poisson-conditioned time discretization scheme} \label{ss:pois-dt}
\noindent
The Poisson conditioning framework also enables the construction of a small time-interval simulation of the integrated variance. The new Poisson-conditioned time discretization scheme competes favorably with \citet{andersen2008simple}'s QE scheme.

\vspace{1ex} \noindent
\textbf{Sampling variance}. To take advantage of Poisson conditioning, we sample $\var_i$ using the Gamma-Poisson scheme in \eqref{eq:var_T}. In the time discretization setup, sampling $\var_{i+1}$ given $\var_i$ is modified to
\begin{equation} \label{eq:var}
\mu_i \sim \POIS\left(\frac{\var_i \phi_\dt(\mr)e^{-\frac{\mr\dt}{2}}}{2}\right), \qtext{then} \var_{i+1} \sim \frac{2e^{-\frac{\mr\dt}{2}}}{\phi_\dt(\mr)}\GAM \left(\frac{\delta}{2}+\mu_i \right),
\end{equation}
where we use the subscript $i$ instead of $t$ by following a convention similar to that in Section~\ref{ss:qe}. 

Since \eqref{eq:var} represents an exact sampling of $\var_{i+1}$, it is more accurate than \citet{andersen2008simple}'s QE scheme for large time steps. The concern is computing speed. Surprisingly, our numerical tests (see Section~\ref{s:num}) verify that the execution time is comparable to that of QE approximation. It may become slower than the QE step when time step $h$ becomes very small. This is because the Poisson rate of $\mu_i$ grows as $2\var_i/(\vov^2 h)$ for a small $h$; therefore, it takes more time to simulate $\mu_i$.

\vspace{1ex}
\noindent
\textbf{Poisson-conditioned quadrature approximation}. Next, we determine $I_{i,i+1}$ given $\var_i$ and $\var_{i+1}$. Given the results in \eqref{eq:iv-pois-mv}, the natural choice for the deterministic value representing $\ivstep$ is its mean:
\begin{equation} \label{eq:pois-td}
\ivstep^\text{POIS} = E(\ivstep\,|\,\mu_i) = (\var_i + \var_{i+1}) m_X \dt + \left(\frac{\delta}{2} + 2\mu_i\right) m_Z \vov^2 \dt^2, \\
\end{equation}
where $m_X$ and $m_Z$ in \eqref{eq:mv} should be redefined with $a=\mr\dt/2$ under the time discretization. As $m_X$ and $m_Z$ assume constant values independent of the simulation path, it is necessary to calculate them only once. Given the simulation path, the conditional integrated variance over the entire time period is given by
\begin{equation} \label{eq:pois-td-sum}
\ivi^\text{POIS} = (\var_0 + 2\var_1 + \cdots + 2\var_{N-1} + \var_N) m_X \dt + \left[\frac{N\delta}{2} + 2(\mu_0 + \mu_1 + \cdots + \mu_{N-1})\right] m_Z \vov^2 \dt^2. 
\end{equation}
Note that this approach is practically feasible owing to the computationally simple expression in \eqref{eq:iv-pois-mv} under Poisson conditioning. The same approach without Poisson conditioning is not feasible since evaluation of the modified Bessel function is required in \eqref{eq:eta-mv} to calculate $E(\ivt)$ and $\text{Var}(\ivt)$ in \eqref{eq:iv-mv}. 

The new approximation $\ivstep^\text{POIS}$ improves over the naive trapezoidal approximation $\ivstep^\text{TZ}$ in \eqref{eq:tz} and forms the building block for our time discretization scheme for pricing path-dependent options. Indeed, $\ivstep^\text{POIS}$ can be shown to be related to $\ivstep^\text{TZ}$ in the limit of $h\downarrow 0$. More precisely, based on the asymptotic expansion in powers of $\dt$ (see \ref{apdx:asym}), it is instructive to observe
\begin{equation} \label{eq:Ihat}
\lim_{\dt\downarrow 0} E\left(\ivstep^\text{POIS}\right) = \left(\var_i + \var_{i+1} + \sqrt{\var_i \var_{i+1}}\right)\frac{\dt}{3}
\approx (\var_i + \var_{i+1})\frac{\dt}{2} = \ivstep^\text{TZ}.
\end{equation}
It is not surprising that the asymptotic form of $\ivstep^\text{POIS}$ involves the arithmetic average, $\frac{\var_i + \var_{i+1}}{2}$, and the geometric average, $\sqrt{\var_i \var_{i+1}}$, as these two quantities also appear in the Laplace transform of $\ivt$ (see \eqref{eq:lap-heston}). Furthermore, the asymptotic expansion of $\ivstep^\text{POIS}$ in \ref{apdx:asym} reveals that the leading order of truncation in the trapezoidal rule is $O(\dt^2)$. This may explain why the trapezoidal rule enjoys good accuracy in approximating $\ivstep$ when the time step $\dt$ is small.

\vspace{1ex}
\noindent
\textbf{Martingale correction}. We derive the corresponding martingale correction in our scheme, which is different from that in the QE scheme. The martingale correction starts by recognizing that $\text{Var}(\ivstep\,|\,\mu_i)$, though small, has been ignored because $\ivstep^\text{POIS}$ is a deterministic value. From \eqref{eq:iv-pois-mv}, the missing variance is given by
$$ \text{Var}(\ivstep\,|\,\mu_i) = (\var_i + \var_{i+1}) v_X \vov^2 \dt^3 + \left(\frac{\delta}{2} + 2\mu_i\right) v_Z \vov^4 \dt^4.
$$
For a random variable $X$, the following approximation holds if the variance is small:
$$ E(e^{a+bX}) \approx e^{a+bE(X) + \frac12 b^2\text{Var}(X)}.
$$
Therefore, under Poisson-conditioned time discretization, the conditional forward price \eqref{eq:cond-fwd} is corrected to 
\begin{equation}\label{eq:fwd-pois}
\begin{aligned}
F_{i+1}|\mu_i &= S_i e^{(r-q)h} E\left(\exp \left(-\frac{\rho^2}{2} \ivstep + \frac{\rho}{\vov}[\var_{i+1} -\var_i + \mr (\ivstep - \vinf T)]\right)\,\Big|\,\mu_i\right)\\
&\approx S_i e^{(r-q)h} \exp \left(-\frac{\rho^2}{2} \ivstep^\text{POIS} + \frac{\rho}{\vov}[\var_{i+1} -\var_i + \mr (\ivstep^\text{POIS} - \vinf T)] + M_{i,i+1}^\text{POIS} \right),
\end{aligned}
\end{equation}
where the martingale correction term $M_{i,i+1}^\text{POIS}$ is given by
\begin{equation} \label{eq:mc-price}
\begin{aligned}
M_{i,i+1}^\text{POIS} &= \frac{\rho^2}2 \left(\frac{\mr}{\vov}-\frac{\rho}{2}\right)^2 \text{Var}(\ivstep|\mu_i)\\
&= \frac{\rho^2}2 \left(\frac{\mr}{\vov}-\frac{\rho}{2}\right)^2\left[(\var_i + \var_{i+1}) v_X + \left(\frac{\delta}{2} + 2\mu_i\right) v_Z \vov^2 \dt\right]\vov^2\dt^3.
\end{aligned}
\end{equation}
Note a subtle difference in the derivation of the martingale correction term when compared with the QE scheme. In the QE scheme, $M_{i,i+1}^\text{QE}$ is fitted to satisfy the unconditional expectation, $S_i = e^{(q-r)h} E(F_{i+1}|S_i)$. In contrast, in our scheme, we use the conditional expectation \eqref{eq:cond-fwd}. Consequently, $M_{i,i+1}^\text{POIS}$ is a function of both $\var_i$ and $\var_{i+1}$ whereas $M_{i,i+1}^\text{QE}$ depends solely on $\var_i$. 

This approach provides more flexibility in determining the correction for missing variance in different contexts. In pricing variance swaps (see Section~\ref{ss:varswap}), we need to sample the realized return variance $\ln^2(S_{i+1}/S_i)$ rather than $S_i$ itself. Using the equality, we have
$$ E\left( (a+bX)^2 \right) = \left(a+bE(X)\right)^2 + b^2\text{Var}(X).
$$
We can correct the realized return variance sampling as
\begin{equation*}
\ln^2(S_{i+1}/S_i) 
\approx \left((r-q)\dt - \frac{\ivstep^\text{POIS}}{2} + \frac{\rho}{\vov} \left[\var_{i+1} - \var_i + \mr(\ivstep^\text{POIS} - \vinf h) \right] + \Sigma_{i,i+1} Z \right)^2 + {M'}_{i,i+1}^\text{POIS},
\end{equation*}
where the correction terms are different. Note that
\begin{equation} \label{eq:mc-return}
{M'}_{i,i+1}^\text{POIS} = \left(\frac{\rho\mr}{\vov}-\frac{1}{2}\right)^2 \text{Var}(\ivstep|\mu_i).
\end{equation}
In the QE scheme, it is unclear how to find martingale correction in the context of the realized return variance. We verify the effectiveness of the new martingale corrections, \eqref{eq:mc-price} and \eqref{eq:mc-return}, for pricing European options and variance swaps, respectively, in our numerical tests in Section~\ref{s:num}.

\section{Numerical performance of the simulation schemes} \label{s:num}
\noindent
We performed comprehensive numerical tests to assess the performance of the proposed simulation schemes with Poisson conditioning in comparison to existing schemes. Specifically, we price European vanilla options in Section~\ref{ss:opt} and variance swaps in Section~\ref{ss:varswap} using various schemes. For easier reference, we label the methods to be tested as follows.
\begin{itemize}
	\item \textbf{GE}: \citet{glasserman2011gamma}'s original GE scheme in Section~\ref{ss:ge}
	\item \textbf{IG}: \citet{tse2013lowbias}'s IG approximation in Section~\ref{ss:ig}.
	\item \textbf{QEM}: \citet{andersen2008simple}'s QE scheme with the trapezoidal rule and martingale correction in Section~\ref{ss:qe}.
	\item \textbf{POIS--GE}: Poisson-conditioned GE scheme in Section~\ref{ss:pois-ge}.
	\item \textbf{POIS--TD}: Poisson-conditioned time discretization scheme in Section~\ref{ss:pois-dt}
\end{itemize}

The methods described above are implemented in Python.
We aim to keep the implementation of the schemes as simple as possible for a succinct and fair performance comparison. In this regard, we do not adopt the IPZ algorithm~\citep[Algorithms 1 and 2]{tse2013lowbias}, which is a technique to speed up the sampling of \eqref{eq:var_T} using the tabulated inverse distribution function of $\GAM(\delta/2)$ corresponding to $\mu_0 = 0$. As the IPZ algorithm would benefit both the existing and proposed methods (except QEM), its implementation is not quite necessary when comparing the performance. Instead, we use the standard random number generation routines available in Python.\footnote{We use \href{https://numpy.org/doc/stable/reference/random/generated/numpy.random.noncentral_chisquare.html}{\texttt{numpy.random.noncentral\_chisquare}} if $\mu_0$ is not required (e.g., IG and GE). We use \href{https://numpy.org/doc/stable/reference/random/generated/numpy.random.poisson.html}{\texttt{numpy.random.poisson}} and \href{https://numpy.org/doc/stable/reference/random/generated/numpy.random.standard_gamma.html}{\texttt{numpy.random.standard\_gamma}} if $\mu_0$ is required (e.g., POIS--GE and POIS--TD).} We do not use the tabulation--interpolation of $E(\eta_{0,T})$ and $\text{Var}(\eta_{0,T})$ in IG. This would speed up IG at the expense of implementation complexity. Such a trick is not necessary in the corresponding POIS--GE ($K=0$). 

\begin{table}[ht]
\begin{center} \small
	\begin{tabular}{|c||c|c|c|c|c|c|c|c||c|c|c|c|} \hline
		Case & $\var_0$ & $\vinf$ & $\vov$ & $\rho$ & $\mr$ & $T$ & $r$ (\%) & $q$ (\%) & $X$ & $C_H$ & $E(R_{0,T})$ & $\text{Var}(R_{0,T})$ \\ \hline\hline
I & 0.04 & 0.04 & 1 & -0.9 & 0.5 & 10 & 0 & 0 & 100 & 13.08467014 & 0.04 & 0.011243 \\
II & 0.04 & 0.04 & 0.9 & -0.5 & 0.3 & 15 & 0 & 0 & 100 & 16.64922292 & 0.04 & 0.016118 \\
III & 0.010201 & 0.019 & 0.61 & -0.7 & 6.21 & 1 & 3.19 & 0 & 100 & ~6.80611331 & 0.017586 & 0.000126 \\
IV & 0.04 & 0.25 & 1 & -0.5 & 4 & 1 & 1 & 2 & 120 & ~9.02491348 & 0.198462 & 0.007109 \\
 \hline
	\end{tabular}
\end{center}
\caption{\label{t:params}The four sets of the Heston model parameters used in numerical tests. All cases assume $S_0=100$. The exact call option price ($C_H$) with strike price $X$, mean and variance of the average variance ($E(R_{0,T})$ and $\text{Var}(R_{0,T})$) are also provided for reference.}
\end{table}

We adopt the four sets of the Heston model parameter values in Table~\ref{t:params}. The first three parameter sets have been frequently used in earlier studies.
\begin{itemize}
  \item Case~I: \citet{andersen2008simple}, \citet{vanhaastrecht2010efficient}, \citet{lord2010comparison}, \citet{tse2013lowbias}
  \item Case~II: \citet{andersen2008simple}, \citet{vanhaastrecht2010efficient}
  \item Case~III: \citet{broadie2006mcheston}, \citet{tse2013lowbias}
\end{itemize}
We add Case~IV~\citep{lewis2019heston} to explore a new case. Unlike the first three cases, Case~IV does not violate the Feller condition (i.e., $\delta=4$) and the strike price is out-of-the-money. Overall, our test cases are fairly diverse. For example, Cases I and II are long-dated options, while Cases III and IV exhibit strong mean reversion ($\mr$). 

Table~\ref{t:params} also reports the exact call option price, and the mean and variance of the average variance for reference. The exact option prices are obtained from the inverse fast Fourier transform~\citep{lewis2000sv} with the unconditional characteristic function of the log asset price, which is free from the branch cut discontinuity~\citep{lord2010complex}. The prices obtained in this approach agree well with the high-precision values reported in the literature.

We present a comparison of efficiency (CPU time) and accuracy (quantified by bias and standard error) of the derivative prices obtained with various simulation schemes. For the simulation estimator $\hat{\Theta}$ and the true price $\Theta$, the bias and standard error (SE) of the estimator are respectively defined by 
$$ \text{Bias} = E(\hat{\Theta}) - \Theta \qtext{and} \text{SE} = \sqrt{E(\hat{\Theta}^2) - E(\hat{\Theta})^2 }.$$
In all the simulation experiments, we drew $160,000$ simulation paths to obtain an estimator $\hat{\Theta}$ and repeated the same experiment 200 times to obtain $E(\hat{\Theta})$ and $E(\hat{\Theta}^2)$.
We ran the simulation experiments on a PC running Windows 11 with an Intel i7--11700 (2.5 GHz) CPU and 8 GB RAM.

\begin{table}[ht]
\begin{center} \small
\begin{tabular}{c|c||c|c|c||c|c|c} \hline
	\multicolumn{2}{c||}{Case I} & \multicolumn{3}{c||}{\textbf{GE}} & \multicolumn{3}{c}{\textbf{POIS--GE}} \\ \hline
	& & Time & Option & Spot & Time & Option & Spot \\
	$N$ & $K$ & (sec) & Bias (SE) & Bias (SE) & (sec) & Bias (SE) & Bias (SE) \\ \hline
1 & 0 & 0.095 & ~2.481 (0.025) & ~1.892 (0.089) & 0.038 & ~0.153 (0.020) & ~0.069 (0.078) \\
1 & 1 & 0.105 & ~0.987 (0.022) & ~0.340 (0.076) & 0.047 & ~0.154 (0.020) & ~0.057 (0.074) \\
1 & 2 & 0.116 & ~0.409 (0.021) & ~0.075 (0.075) & 0.056 & ~0.084 (0.019) & ~0.014 (0.074) \\
1 & 4 & 0.130 & ~0.087 (0.020) & ~0.001 (0.078) & 0.074 & ~0.023 (0.019) & -0.000 (0.074) \\
1 & 8 & 0.163 & ~0.006 (0.019) & -0.000 (0.076) & 0.108 & ~0.002 (0.019) & -0.003 (0.077) \\
	\hline
\end{tabular} \vspace{1em}

\begin{tabular}{c|c||c|c|c||c|c|c} \hline
	\multicolumn{2}{c||}{Case I} & \multicolumn{3}{c||}{\textbf{IG}} & \multicolumn{3}{c}{\textbf{POIS--GE ($K=0$)}} \\ \hline
	& & Time & Option & Spot & Time & Option & Spot \\
	$N$ & $\dt$ & (sec) & Bias (SE) & Bias (SE) & (sec) & Bias (SE) & Bias (SE) \\ \hline
1 & 10 & 0.097 & ~0.159 (0.019) & ~0.093 (0.077) & 0.038 & ~0.153 (0.020) & ~0.069 (0.078) \\
2 & 5 & 0.168 & -0.057 (0.020) & -0.205 (0.075) & 0.057 & -0.057 (0.020) & -0.160 (0.076) \\
4 & 2.5 & 0.308 & -0.136 (0.020) & -0.119 (0.081) & 0.095 & -0.105 (0.019) & -0.055 (0.076) \\
8 & 1.25 & 0.593 & -0.068 (0.018) & -0.023 (0.077) & 0.172 & -0.043 (0.020) & -0.014 (0.075) \\
	\hline
\end{tabular} \vspace{1em}

\begin{tabular}{c|c||c|c|c||c|c|c} \hline
	\multicolumn{2}{c||}{Case I} & \multicolumn{3}{c||}{\textbf{QEM}} & \multicolumn{3}{c}{\textbf{POIS--TD}} \\ \hline
	& & Time & Option & Spot & Time & Option & Spot \\
	$N$ & $\dt$ & (sec) & Bias (SE) & Bias (SE) & (sec) & Bias (SE) & Bias (SE) \\ \hline
20 & 1/2 & 0.381 & ~0.116 (0.021) & -0.011 (0.082) & 0.266 & -0.115 (0.019) & ~0.003 (0.071) \\
40 & 1/4 & 0.733 & ~0.008 (0.019) & -0.011 (0.084) & 0.527 & -0.030 (0.020) & -0.009 (0.078) \\
80 & 1/8 & 1.505 & -0.015 (0.019) & -0.015 (0.078) & 1.100 & -0.004 (0.020) & ~0.015 (0.076) \\
	\hline
\end{tabular}
\end{center}
\caption{\label{t:opt1}Speed-accuracy comparison of the existing (left) and Poisson-conditioned (right) schemes for Case~I. The bias and standard error of the at-the-money ($X=S_0=100$) call option in \eqref{eq:C} and the spot price in \eqref{eq:S0} are reported. See Table~\ref{t:params} for the parameter values and reference option price. For the GE schemes (top), we take $N=1$ time step ($h=T$) while the number of gamma terms, $K$, is varied. For the IG approximations (middle) and time discretization schemes (bottom), $N$ and $h$ are varied, respectively. }
\end{table}

\begin{table}[ht]
\begin{center} \small
\begin{tabular}{c|c||c|c|c||c|c|c} \hline
	\multicolumn{2}{c||}{Case II} & \multicolumn{3}{c||}{\textbf{GE}} & \multicolumn{3}{c}{\textbf{POIS--GE}} \\ \hline
	& & Time & Option & Spot & Time & Option & Spot \\
	$N$ & $K$ & (sec) & Bias (SE) & Bias (SE) & (sec) & Bias (SE) & Bias (SE) \\ \hline
1 & 0 & 0.096 & -1.950 (0.011) & ~0.686 (0.062) & 0.037 & -0.107 (0.011) & -0.003 (0.054) \\
1 & 1 & 0.104 & -0.823 (0.013) & ~0.088 (0.055) & 0.044 & -0.121 (0.011) & ~0.014 (0.051) \\
1 & 2 & 0.113 & -0.366 (0.012) & ~0.014 (0.056) & 0.052 & -0.075 (0.010) & ~0.006 (0.055) \\
1 & 4 & 0.128 & -0.089 (0.012) & -0.004 (0.058) & 0.071 & -0.026 (0.012) & -0.006 (0.052) \\
1 & 8 & 0.164 & -0.006 (0.011) & -0.003 (0.054) & 0.104 & -0.003 (0.012) & -0.005 (0.054) \\
	\hline
\end{tabular} \vspace{1em}

\begin{tabular}{c|c||c|c|c||c|c|c} \hline
	\multicolumn{2}{c||}{Case II} & \multicolumn{3}{c||}{\textbf{IG}} & \multicolumn{3}{c}{\textbf{POIS--GE ($K=0$)}} \\ \hline
	& & Time & Option & Spot & Time & Option & Spot \\
	$N$ & $\dt$ & (sec) & Bias (SE) & Bias (SE) & (sec) & Bias (SE) & Bias (SE) \\ \hline
1 & 15 & 0.089 & -0.117 (0.011) & ~0.006 (0.056) & 0.037 & -0.107 (0.011) & -0.003 (0.054) \\
2 & 7.5 & 0.156 & ~0.064 (0.011) & -0.103 (0.058) & 0.056 & ~0.065 (0.010) & -0.071 (0.056) \\
4 & 3.75 & 0.297 & ~0.122 (0.010) & -0.034 (0.055) & 0.096 & ~0.096 (0.010) & -0.014 (0.053) \\
8 & 1.875 & 0.582 & ~0.066 (0.009) & ~0.003 (0.052) & 0.174 & ~0.044 (0.011) & ~0.000 (0.055) \\
	\hline
\end{tabular} \vspace{1em}

\begin{tabular}{c|c||c|c|c||c|c|c} \hline
	\multicolumn{2}{c||}{Case II} & \multicolumn{3}{c||}{\textbf{QEM}} & \multicolumn{3}{c}{\textbf{POIS--TD}} \\ \hline
	& & Time & Option & Spot & Time & Option & Spot \\
	$N$ & $\dt$ & (sec) & Bias (SE) & Bias (SE) & (sec) & Bias (SE) & Bias (SE) \\ \hline
30 & 1/2 & 0.568 & -0.124 (0.011) & -0.004 (0.057) & 0.390 & ~0.078 (0.009) & -0.001 (0.055) \\
60 & 1/4 & 1.119 & -0.010 (0.010) & -0.006 (0.060) & 0.755 & ~0.017 (0.010) & ~0.000 (0.053) \\
120 & 1/8 & 2.178 & ~0.009 (0.010) & -0.004 (0.055) & 1.540 & ~0.005 (0.010) & ~0.001 (0.054) \\
	\hline
\end{tabular}

\end{center}
\caption{\label{t:opt2}Speed-accuracy comparison of the existing (left) and Poisson-conditioned (right) schemes for Case~II. The bias and standard error of the at-the-money ($X=S_0=100$) call option in \eqref{eq:C} and the spot price in \eqref{eq:S0} are reported. See Table~\ref{t:params} for the parameter values and reference option price. For the GE schemes (top), we take $N=1$ time step ($h=T$) while the number of gamma terms, $K$, is varied. For the IG approximations (middle) and time discretization schemes (bottom), $N$ and $h$ are varied, respectively.}
\end{table}

\begin{table}[ht]
\begin{center} \small
\begin{tabular}{c|c||c|c|c||c|c|c} \hline
	\multicolumn{2}{c||}{Case III} & \multicolumn{3}{c||}{\textbf{GE}} & \multicolumn{3}{c}{\textbf{POIS--GE}} \\ \hline
	& & Time & Option & Spot & Time & Option & Spot \\
	$N$ & $K$ & (sec) & Bias (SE) & Bias (SE) & (sec) & Bias (SE) & Bias (SE) \\ \hline
1 & 0 & 0.097 & ~0.078 (0.011) & ~0.006 (0.022) & 0.042 & ~0.005 (0.011) & -0.000 (0.025) \\
1 & 1 & 0.107 & ~0.008 (0.010) & ~0.001 (0.023) & 0.053 & ~0.001 (0.010) & ~0.000 (0.022) \\
1 & 2 & 0.118 & ~0.001 (0.011) & -0.000 (0.024) & 0.063 & ~0.001 (0.009) & ~0.001 (0.022) \\
1 & 4 & 0.142 & ~0.001 (0.010) & ~0.001 (0.022) & 0.084 & -0.000 (0.010) & -0.001 (0.022) \\
1 & 8 & 0.193 & ~0.000 (0.010) & ~0.000 (0.024) & 0.131 & -0.000 (0.011) & -0.001 (0.024) \\
	\hline
\end{tabular} \vspace{1em}

\begin{tabular}{c|c||c|c|c||c|c|c} \hline
	\multicolumn{2}{c||}{Case III} & \multicolumn{3}{c||}{\textbf{IG}} & \multicolumn{3}{c}{\textbf{POIS--GE ($K=0$)}} \\ \hline
	& & Time & Option & Spot & Time & Option & Spot \\
	$N$ & $\dt$ & (sec) & Bias (SE) & Bias (SE) & (sec) & Bias (SE) & Bias (SE) \\ \hline
1 & 1 & 0.106 & ~0.007 (0.010) & ~0.001 (0.024) & 0.042 & ~0.005 (0.011) & -0.000 (0.025) \\
2 & 1/2 & 0.213 & -0.008 (0.011) & ~0.001 (0.023) & 0.064 & -0.006 (0.011) & ~0.001 (0.023) \\
4 & 1/4 & 0.444 & -0.005 (0.009) & -0.000 (0.020) & 0.105 & -0.001 (0.011) & ~0.000 (0.023) \\
8 & 1/8 & 1.025 & -0.001 (0.009) & -0.002 (0.020) & 0.192 & ~0.001 (0.009) & ~0.001 (0.021) \\
	\hline
\end{tabular} \vspace{1em}

\begin{tabular}{c|c||c|c|c||c|c|c} \hline
	\multicolumn{2}{c||}{Case III} & \multicolumn{3}{c||}{\textbf{QEM}} & \multicolumn{3}{c}{\textbf{POIS--TD}} \\ \hline
	& & Time & Option & Spot & Time & Option & Spot \\
	$N$ & $\dt$ & (sec) & Bias (SE) & Bias (SE) & (sec) & Bias (SE) & Bias (SE) \\ \hline
2 & 1/2 & 0.061 & ~0.097 (0.005) & -0.001 (0.016) & 0.046 & -0.467 (0.008) & ~0.003 (0.018) \\
4 & 1/4 & 0.098 & ~0.013 (0.009) & -0.002 (0.022) & 0.075 & -0.164 (0.010) & -0.000 (0.021) \\
8 & 1/8 & 0.170 & -0.009 (0.010) & -0.001 (0.022) & 0.133 & -0.045 (0.010) & ~0.000 (0.021) \\
	\hline
\end{tabular}

\end{center}
	\caption{\label{t:opt3}Speed-accuracy comparison of the existing (left) and Poisson-conditioned (right) schemes for Case~III. The bias and standard error of the at-the-money ($X=S_0=100$) call option in \eqref{eq:C} and the spot price in \eqref{eq:S0} are reported. See Table~\ref{t:params} for the parameter values and reference option price. For the GE schemes (top), we take $N=1$ time step ($h=T$) while the number of gamma terms, $K$, is varied. For the IG approximations (middle) and time discretization schemes (bottom), $N$ and $h$ are varied, respectively.}
\end{table}

\begin{table}[ht]
\begin{center} \small
\begin{tabular}{c|c||c|c|c||c|c|c} \hline
	\multicolumn{2}{c||}{Case IV} & \multicolumn{3}{c||}{\textbf{GE}} & \multicolumn{3}{c}{\textbf{POIS--GE}} \\ \hline
	& & Time & Option & Spot & Time & Option & Spot \\
	$N$ & $K$ & (sec) & Bias (SE) & Bias (SE) & (sec) & Bias (SE) & Bias (SE) \\ \hline
1 & 0 & 0.084 & ~0.032 (0.013) & ~0.015 (0.049) & 0.035 & -0.001 (0.013) & ~0.000 (0.053) \\
1 & 1 & 0.092 & ~0.002 (0.012) & ~0.000 (0.048) & 0.043 & ~0.000 (0.013) & ~0.002 (0.052) \\
1 & 2 & 0.101 & -0.000 (0.012) & -0.001 (0.050) & 0.053 & ~0.001 (0.013) & ~0.002 (0.054) \\
1 & 4 & 0.124 & -0.001 (0.013) & -0.002 (0.054) & 0.073 & -0.000 (0.013) & -0.001 (0.049) \\
1 & 8 & 0.170 & ~0.000 (0.013) & ~0.002 (0.052) & 0.107 & ~0.000 (0.013) & ~0.000 (0.053) \\
	\hline
\end{tabular} \vspace{1em}

\begin{tabular}{c|c||c|c|c||c|c|c} \hline
	\multicolumn{2}{c||}{Case IV} & \multicolumn{3}{c||}{\textbf{IG}} & \multicolumn{3}{c}{\textbf{POIS--GE ($K=0$)}} \\ \hline
	& & Time & Option & Spot & Time & Option & Spot \\
	$N$ & $\dt$ & (sec) & Bias (SE) & Bias (SE) & (sec) & Bias (SE) & Bias (SE) \\ \hline
1 & 1 & 0.091 & -0.001 (0.012) & -0.002 (0.051) & 0.035 & -0.001 (0.013) & ~0.000 (0.053) \\
2 & 1/2 & 0.223 & -0.004 (0.012) & -0.003 (0.049) & 0.052 & -0.001 (0.013) & ~0.005 (0.053) \\
4 & 1/4 & 0.609 & -0.001 (0.012) & -0.000 (0.050) & 0.092 & -0.002 (0.014) & -0.004 (0.055) \\
8 & 1/8 & 1.299 & -0.000 (0.013) & -0.001 (0.051) & 0.172 & ~0.002 (0.013) & ~0.007 (0.053) \\
	\hline
\end{tabular} \vspace{1em}

\begin{tabular}{c|c||c|c|c||c|c|c} \hline
	\multicolumn{2}{c||}{Case IV} & \multicolumn{3}{c||}{\textbf{QEM}} & \multicolumn{3}{c}{\textbf{POIS--TD}} \\ \hline
	& & Time & Option & Spot & Time & Option & Spot \\
	$N$ & $\dt$ & (sec) & Bias (SE) & Bias (SE) & (sec) & Bias (SE) & Bias (SE) \\ \hline
2 & 1/2 & 0.047 & -0.599 (0.005) & -0.001 (0.014) & 0.039 & -0.096 (0.012) & ~0.013 (0.049) \\
4 & 1/4 & 0.071 & -0.166 (0.005) & -0.001 (0.016) & 0.063 & -0.034 (0.013) & -0.003 (0.053) \\
8 & 1/8 & 0.129 & -0.045 (0.005) & -0.002 (0.016) & 0.113 & -0.007 (0.013) & ~0.008 (0.052) \\
	\hline
\end{tabular}
\end{center}
\caption{\label{t:opt4}Speed-accuracy comparison of the existing (left) and Poisson-conditioned (right) schemes for Case~IV. The bias and standard error of the out-of-the-money ($X=120$) call option in \eqref{eq:C} and the spot price in \eqref{eq:S0} are reported. See Table~\ref{t:params} for the parameter values and reference option price. For the GE schemes (top), we take $N=1$ time step ($h=T$) while the number of gamma terms, $K$, is varied. For the IG approximations (middle) and time discretization schemes (bottom), $N$ and $h$ are varied, respectively.}
\end{table}

\subsection{European call options} \label{ss:opt}
\noindent 
For pricing the European vanilla option, we use the conditional MC method~\citep{willard1997condmc} instead of simulating $S_T$. Conditional on $\var_T$ and $\ivt$, $S_T$ follows a geometric Brownian motion in \eqref{eq:gbm}, and the option price can be obtained using the BS formula with forward price $F_T$ in \eqref{eq:cond-fwd} and volatility $\sigma = \Sigma_{0,T}/\sqrt{T}$. 
Therefore, the unconditional European call option price under the Heston model can be estimated by taking the MC average of the BS prices over the simulated values of $\var_T$ and $\ivt$:
\begin{equation}\label{eq:C}
\hat{C}_H = e^{-rT}\, E_\text{MC}\{C_\text{BS} (F_T, \sigma, T, X)\},
\end{equation}
where $C_\text{BS}$ is the undiscounted BS call option price, with forward price $F_T$, volatility $\sigma$, maturity $T$, and strike price $X$.\footnote{The option Greeks can also be obtained by replacing the price $C_\text{BS}$ in \eqref{eq:C} with the relevant BS Greek formula.} As the MC variance from the sampling of $S_T$ is suppressed by the BS formula, the conditional MC reduces the MC variance of $\hat{C}_H$, thereby increasing the accuracy of the bias of the simulation methods. This procedure has been used in \citet[Tables~4 and 5]{broadie2006mcheston} for the Heston model
and \citet{cai2017sabr} for the SABR model.

We also measure how accurately the martingale condition is preserved. In theory, the unconditional expectation of $F_T$ should equal the forward price, $e^{(r-q)T}S_0$. Therefore, we reconstruct the spot price by taking the discounted MC average of $F_T$:
\begin{equation}\label{eq:S0}
\hat{S_0} = e^{(q-r)T}\, E_\text{MC}\{F_T\},
\end{equation}
and examines the extent to which $\hat{S_0}$ differs from $S_0$. This serves as another measure to assess the accuracy of our proposed simulation schemes, in particular, the effectiveness of the martingale correction in POIS--TD. In the two time discretization schemes, we apply the aggregate martingale correction term to the conditional forward $F_T$:
$$ M_{0,N} = \sum_{i=0}^{N-1} M_{i,i+1}. 
$$
In POIS--TD, $M_{0,N}^\text{POIS}$ is proportional to
\begin{equation} \label{eq:var_sum}
\begin{aligned}
\sum_{i=0}^{N-1}&\text{Var}(\ivstep|\mu_i) = \\
&\left[(\var_0 + 2\var_1 + \cdots + 2\var_{N-1} + \var_N) v_X + \left(\frac{N\delta}{2} + 2\mu_0 + \cdots + 2\mu_{N-1}\right) v_Z \vov^2 \dt\right]\vov^2 \dt^3,
\end{aligned}
\end{equation}
which is a linear combination of the sums of $\var_i$ and $\mu_i$. This adds little extra computation, since the same terms are already used in $\ivi^\text{POIS}$ in \eqref{eq:pois-td-sum}.

Tables~\ref{t:opt1}--\ref{t:opt4} show the numerical performance of the four cases. In each table, we compare the GE-based exact schemes (top), IG-based low-bias schemes (middle), and time discretization schemes (bottom). We comment on each of these comparisons below.

In the GE versus POIS--GE comparison, POIS--GE shows substantially lower bias with CPU time reduction (typically 40\%). The IG approximation contributes to the lower bias for the truncated series, and Poisson conditioning contributes to the faster execution. For POIS--GE with low $K$ values, the biases in Cases I and II are relatively larger than those in the other cases because of the long maturity. Nevertheless, the spot price is accurately preserved, indicating that the distributional error in $\ivt$ due to low $K$ manifests in option prices rather than spot prices. The option price bias becomes exceedingly small when $K$ increases to 8. 

In the IG versus POIS--GE ($K=0$) comparison, the latter reduces the computation time by several factors, though the bias improvement is marginal. This is because we have avoided evaluating the modified Bessel function present in the original IG scheme.

In the QEM versus POIS--TD comparison, POIS--TD compares favorably with QEM. The use of better discretization rules leads to an improved accuracy. It is also noteworthy that the runtime efficiency of POIS--TD is slightly better than that of the QEM. This is surprising given that the motivation for introducing the QE step \eqref{eq:qe} is to avoid directly drawing a noncentral chi-squared variate. As expected in the time discretization schemes, both schemes require a small $h$ to achieve high accuracy. Despite the short maturity, Cases III and IV show relatively larger biases than the other cases for a large time step $h$. However, the spot price bias is close to zero even with a large $h$, which verifies that the martingale correction is effective in both schemes. The advantage of POIS--TD is more pronounced for pricing variance swap, more details are discussed in the next subsection.

\begin{table}[ht]
\begin{center} \small
	\begin{tabular}{c|c||cc|cc||cc|cc||cc|cc} \hline
		\multicolumn{2}{c||}{Case IV} & \multicolumn{2}{c|}{\textbf{GE}} & \multicolumn{2}{c||}{\textbf{POIS--GE}} & \multicolumn{2}{c|}{\textbf{IG}} & \multicolumn{2}{c||}{\textbf{POIS--GE}} & \multicolumn{2}{c|}{\textbf{QEM}} & \multicolumn{2}{c}{\textbf{POIS--TD}} \\ \hline
		$\vov$ & $\mr$ & Time & Bias & Time & Bias & Time & Bias & Time & Bias & Time & Bias & Time & Bias \\ \hline
 & 4 & 0.131 & 0.040 & 0.078 & 0.041 & 0.198 & 0.042 & 0.090 & 0.042 & 0.115 & 0.406 & 0.106 & 0.180 \\
1 & 1 & 0.163 & 0.018 & 0.084 & 0.017 & 0.281 & 0.025 & 0.098 & 0.019 & 0.149 & 0.280 & 0.120 & 0.131 \\
& 0.1 & 0.187 & 0.059 & 0.082 & 0.018 & 0.318 & 0.062 & 0.096 & 0.045 & 0.135 & 0.170 & 0.108 & 0.139 \\ \hline
& 4 & 0.140 & 0.060 & 0.083 & 0.066 & 0.512 & 0.064 & 0.098 & 0.064 & 0.113 & 0.344 & 0.117 & 0.316 \\
0.25 & 1 & 0.165 & 0.034 & 0.085 & 0.039 & 0.645 & 0.036 & 0.096 & 0.038 & 0.111 & 0.045 & 0.117 & 0.036 \\
& 0.1 & 0.258 & 0.012 & 0.084 & 0.012 & 0.635 & 0.012 & 0.096 & 0.012 & 0.111 & 0.007 & 0.113 & 0.014 \\ \hline
& 4 & 0.162 & 0.065 & 0.082 & 0.070 & 0.450 & 0.070 & 0.099 & 0.073 & 0.111 & 0.345 & 0.115 & 0.332 \\
0.1 & 1 & 0.284 & 0.040 & 0.085 & 0.043 & 0.862 & 0.044 & 0.104 & 0.048 & 0.110 & 0.048 & 0.114 & 0.044 \\
& 0.1 & 0.276 & 0.019 & 0.086 & 0.020 & 0.338 & 0.019 & 0.104 & 0.020 & 0.111 & 0.007 & 0.116 & 0.019 \\
		\hline
	\end{tabular} \vspace{1em}
\end{center}
\caption{\label{t:opt4grid}Speed-accuracy comparison of the existing and Poisson-conditioned schemes for Case~IV for a grid of vol-of-vol, $\vov\in\{1, 0.25, 0.1\}$, and mean reversion $\mr\in\{4, 1, 0.1\}$. The bias is measured as the sum of the absolute error in call option prices in \eqref{eq:C} for $K=100,\,110$ and 120.
We use $N=K=1$ in the first comparison (\textbf{GE} versus \textbf{POIS--GE}), 
$h=1/2$ ($N=2$) with $K=0$ in the second comparison (\textbf{IG} versus \textbf{POIS--GE}), and $h=1/4$ ($N=4$) in the third comparison (\textbf{QEM} versus \textbf{POIS--TD}).}
\end{table}

Lastly, we examine whether the advantages of our new simulation scheme remain valid over a wide range of parameter values. Table~\ref{t:opt4grid} shows the computation time and option price bias when the vol-of-vol $\vov$ and mean reversion $\mr$ are varied from the base values of Case IV. We confirm that the Poisson-conditioned methods give performance superior to the existing methods; POIS--GE shows its advantage in computation time while POIS--TD shows its advantage in accuracy.

\subsection{Discretely monitored variance swap} \label{ss:varswap}
\noindent
In general, a swap product is a contractual agreement to exchange the floating and fixed legs of payment for future time $t=T$. In the variance swap, the floating leg is given by the annualized log return variance over $N = T/\dt$ monitoring periods, which is proxied by
\begin{equation}\label{eq:varswap-dt}
  R_{0,T}^\dt = \frac{1}{T} \sum_{i=1}^{N} \ln^2(S_i/S_{i-1}).
\end{equation}
The monitoring frequency is set by the time step $h$: $h=1/4$ for quarterly, $h=1/12$ for monthly, and $h=1/52$ for weekly monitoring.
The value of the fixed leg, $K_\text{swap}^\dt$, is typically determined to ensure that the swap has zero net value at $t=0$:
$$ K_\text{swap}^\dt = E(R_{0,T}^\dt). $$
This value is called the fair strike of the variance swap. The level of path dependence of these variance derivatives goes beyond that of the Asian options since their payoff structures involve the squared sum of the log return of asset price $S_t$ at $t=t_i$. 

The annualized log return variance in the continuous limit ($h \downarrow 0$) is the average variance in \eqref{eq:R}. Note that
\begin{equation}\label{eq:limit}
  \lim_{\dt \downarrow 0} R_{0,T}^\dt \to R_{0,T} = \frac{1}{T} \int_0^T\var_t \;\text{d} t,
\end{equation}
and the continuously monitored fair strike $K_\text{swap}$ is given by \eqref{eq:R_m}:
$$ K_\text{swap} = E(R) = \vinf + (\var_0 - \vinf) \frac{1-e^{- \mr T}}{\mr T}.
$$
The fair strike $K_\text{swap}^\dt$ of the discrete variance swap with the monitoring time step $h$ admits a closed-form solution~\citep{bernard2014prices,kwok2022variance}. We express the discretely monitored fair strike $K_\text{swap}^\dt$ as an adjustment $\Delta_\text{swap}^\dt$ to the continuously monitored fair strike $K_\text{swap}$:
\begin{equation}\label{eq:Kswap}
K_\text{swap}^\dt = E(R_{0,T}^\dt) = K_\text{swap} + \Delta_\text{swap}^\dt,
\end{equation}
where
\begin{align*}
\Delta_\text{swap}^\dt &= \frac{\dt(\vinf + 2q - 2r)}{4} \left[ (\vinf + 2q - 2r) + 2(\var_0 - \vinf) \frac{1-e^{- \mr T}}{\mr T} \right]\\
&+  \frac{\vinf\vov}{\mr} \left( \frac{\vov}{4 \mr} - \rho \right) \left(1 - \frac{1-e^{- \mr\dt}}{\mr\dt} \right) + (\var_0 - \vinf)\frac{\vov}{\mr} \left( \frac{\vov}{2 \mr} - \rho \right) \frac{1-e^{- \mr T}}{\mr T} \left( 1+ \frac{\mr\dt}{1-e^{\mr\dt}} \right)\\
&+ \left[\frac{\vov^2}{\mr^2} (\vinf - 2 \var_0) + \frac{2}{\mr} (\var_0 - \vinf)^2 \right] \frac{1-e^{-2 \mr T}}{8 \mr T} \frac{1-e^{-\mr\dt}}{1+e^{- \mr\dt}}.
\end{align*}
The above analytical solution serves as a benchmark to assess the accuracy of the time discretization schemes. 

We compare the fair strike $K_\text{swap}^\dt$ for various monitoring frequencies ranging from semi-annual ($h=1/2$) to weekly ($h=1/52$). As the time discretization methods are more efficient in pricing derivatives with higher monitoring frequency, such as the variance swap, we only compare QEM with POIS--TD. In POIS--TD, we apply the martingale correction ${M'}_{i,i+1}^\text{POIS}$ in \eqref{eq:mc-return}, which is tailored to log return variance. Given the absence of such a correction in the QEM, we simply use $M_{i,i+1}^\text{QE}$ for the log return. 

\begin{table}[ht]
\begin{center} \small
\begin{tabular}{c|c|c||c|c||c|c} \hline
	\multicolumn{3}{c||}{Case III} & \multicolumn{2}{c||}{\textbf{QEM}} & \multicolumn{2}{c}{\textbf{POIS--TD}} \\ \hline
	& & Benchmark & Time & Bias (SE) & Time & Bias (SE) \\
	$N$ & $\dt$ & $(\times 10^{-2})$ & (sec) & $(\times 10^{-2})$ & (sec) & $(\times 10^{-2})$ \\ \hline
~2 & 1/2 & 1.870 & 0.042 & ~0.041 (0.010) & 0.033 & ~0.000 (0.007) \\
~4 & 1/4 & 1.832 & 0.091 & -0.024 (0.007) & 0.067 & ~0.001 (0.007) \\
12 & 1/12 & 1.790 & 0.241 & -0.011 (0.005) & 0.221 & -0.001 (0.004) \\
52 & 1/52 & 1.767 & 0.954 & -0.000 (0.003) & 0.920 & ~0.000 (0.004) \\
	\hline
\end{tabular}
\end{center}
\caption{\label{t:var3}Speed-accuracy comparison of the time discretization schemes for pricing variance swap with Case~III. The bias and standard error of the fair strike of discretely monitored variance swap are reported for varying monitoring frequencies. The benchmark fair strike is from the analytical reference price in \eqref{eq:Kswap}.}
\end{table}

\begin{table}[ht]
\begin{center} \small
\begin{tabular}{c|c|c||c|c||c|c} \hline
	\multicolumn{3}{c||}{Case IV} & \multicolumn{2}{c||}{\textbf{QEM}} & \multicolumn{2}{c}{\textbf{POIS--TD}} \\ \hline
	& & Benchmark & Time & Bias (SE) & Time & Bias (SE) \\
	$N$ & $\dt$ & $(\times 10^{-2})$ & (sec) & $(\times 10^{-2})$ & (sec) & $(\times 10^{-2})$ \\ \hline
~2 & 1/2 & 21.930 & 0.038 & -0.750 (0.083) & 0.026 & ~0.002 (0.085) \\
~4 & 1/4 & 21.132 & 0.062 & -0.325 (0.060) & 0.057 & ~0.004 (0.063) \\
12 & 1/12 & 20.356 & 0.184 & -0.057 (0.036) & 0.184 & -0.003 (0.038) \\
52 & 1/52 & 19.973 & 0.840 & -0.000 (0.021) & 0.962 & ~0.001 (0.029) \\
	\hline
\end{tabular}
\end{center}
\caption{\label{t:var4}Speed-accuracy comparison of the time discretization schemes for pricing variance swap with Case~IV. The bias and standard error of the fair strike of discretely monitored variance swap are reported for varying monitoring frequencies. The benchmark fair strike is from the analytical reference price in \eqref{eq:Kswap}.}
\end{table}

Tables~\ref{t:var3} and \ref{t:var4} list the results for Cases III and IV, respectively.
The two cases are chosen because variance swaps traded in the market are typically short-dated, such as one year in these cases. They show a relatively larger bias in pricing vanilla options in Section~\ref{ss:opt}. The results show that the POIS--TD bias is much smaller than the QEM bias, though the biases from both schemes quickly converge to zero as the monitoring becomes more frequent. The numerical results also verify that martingale correction \eqref{eq:mc-return} is effective. 

\section{Extension to the multifactor Heston model} \label{s:multi} \noindent
The Poisson-conditioned schemes can be extended to the multifactor Heston model. We extend the classical Heston model, \eqref{eq:hestona}--\eqref{eq:hestonb}, to a multifactor version where $M$ independent stochastic variances drive the price process: 
\begin{align}
	\frac{\text{d} S_t}{S_t} &= (r-q) \;\text{d}t + \sum_{m=1}^M\sqrt{\var_t^m} \left(\rho_m \;\text{d} Z_t^m + \sqrt{1-\rho_m^2} \;\text{d} W_t^m \right),\\
	\text{d} \var_t^m &= \mr_m (\vinf_m - \var_t^m) \;\text{d}t + \vov_m \sqrt{\var_t^m} \;\text{d} Z_t^m,
\end{align}
where $m=1,2,\ldots,M$ and the Brownian motions $\{Z_t^m\}$ and $\{W_t^m\}$ are all independent. The variance factor $\var_t^m$ is correlated to $S_t$ by the correlation coefficient $\rho_m$. This specification of the multifactor Heston model has been popularly adopted in the literature~\citep{christoffersen2009shape,trolle2009generala,recchioni2021complete}.

Suppose the conditional integrated variance is defined for each variance factor by
\begin{equation*}
	\ivtm(\var_0^m, \var_T^m) = \left(\int_0^T \var_t^m\, \text{d}t \;\Big|\; \var_0^m, \var_T^m\right), \quad m=1,2,\ldots,M,
\end{equation*}
the result in \eqref{eq:ln} for the log return sampling is extended to 
\begin{equation} 
	\ln \frac{S_T}{S_0} \sim (r-q)T - \sum_{m=1}^M\frac{\ivtm}{2} + \sum_{m=1}^M\frac{\rho_m}{\vov_m} \left[\var_T^m - \var_0^m + \mr_m(\ivtm - \vinf_m T) \right] + \Sigma_{0,T}^M\; Z,
\end{equation}
where $Z$ is a standard normal variate and $\Sigma_{0,T}^M$ is the total standard deviation as defined by
$$
\Sigma_{0,T}^M = \sqrt{\sum_{m=1}^M(1-\rho_m^2) \ivtm}\;.
$$
As a result, the terminal asset price $S_T$ can be simulated as a geometric Brownian motion, in a similar way to \eqref{eq:gbm} in the classical Heston model:
\begin{equation*} 
	S_T = F_T \exp\left(\Sigma_{0,T}^M\, Z - \frac12 \left(\Sigma_{0,T}^M\right)^2 \right),
\end{equation*}
where $F_T$ is the forward stock price, conditional on $S_0$, $\{\var_0^m\}$, $\{\var_T^m\}$, and $\{\ivtm\}$:
\begin{equation*} 
	\begin{aligned}
		F_T &= E(S_T \,|\,S_0, \var_0^m, \var_T^m, \ivtm)\\
		&= S_0 e^{(r-q)T} \exp \left(-\sum_{i=1}^n\frac{\rho_m^2}{2} \ivtm + \sum_{i=1}^n\frac{\rho_m}{\vov_m}[\var_T^m -\var_0^m + \mr_m (\ivtm - \vinf_m T)]\right).
	\end{aligned}
\end{equation*}
Since the variance factors, $\{\var_t^m\}$, are independent, we can sample $\var_T^m$ and $\ivtm$ for each $m$ according to the Poisson-conditioned GE scheme in Section~\ref{ss:pois-ge}.

\section{Conclusion} \label{s:con} \noindent
Monte Carlo simulation under the Heston model has been a widely studied topic, and several approaches are available for different monitoring frequencies: exact~\citep{broadie2006mcheston,glasserman2011gamma}, low-bias~\citep{tse2013lowbias}, and time discretization~\citep{andersen2008simple} schemes. Among the existing simulation schemes, however, they suffer from computationally expensive evaluations of the modified Bessel function and/or Bessel random variables arising from the square-root variance process. Based on the observation that the conditional integrated variance can be simplified when conditioned by the Poisson variate used for simulating the terminal variance, we propose new simulation methods that enhance the existing methods in all spectra. 

Poisson-conditioned GE is an exact simulation scheme that enhances \citet{glasserman2011gamma}'s GE scheme. It achieves significant speedup by expressing the conditional integrated variance without the Bessel random variable, which represents the computational bottleneck. The adoption of the IG approximation~\citep{tse2013lowbias} for the truncation approximation improves numerical accuracy. A special case of the Poisson-conditioned GE scheme is reduced naturally to \citet{tse2013lowbias}'s low-bias scheme but without the Bessel function. As the Laplace transform of the conditional integrated variance can also be expressed without the Bessel function, \citet{broadie2006mcheston}'s scheme can be expedited. The Poisson-conditioned time discretization scheme is proposed as an alternative to \citet{andersen2008simple}'s QE scheme. Our comprehensive numerical tests illustrate the strong competitiveness of our schemes in speed-accuracy comparison among existing schemes for pricing derivatives under the Heston model. In addition to numerical efficiency, our new Heston simulation schemes are simple and straightforward for practitioners to implement since they involve only the elementary functions and random variables.

We conclude this paper with potential applications of our findings beyond the stochastic volatility model. \citet{glasserman2011gamma} already mentioned that the simulation schemes for the integrated variance can be used in the interest rate term-structure model~\citep{cox1985cir} and stochastic default intensity model. Our simulation methods with Poisson conditioning can further improve speed and accuracy in these applications as well. In particular, \citet{dassios2017efficient} criticize the GE method for its truncation error. As the IG approximation approximates the truncation accurately, we expect that the Poisson-conditioned GE can also be useful in the credit intensity simulation.

\appendix\normalsize
\section{Asymptotic expansion of $\ivstep^\text{POIS}$} \label{apdx:asym}
\noindent
Taking the limit of $h\downarrow 0$, the coefficients in \eqref{eq:mv} have the following asymptotic expansions in powers of $a$:
\begin{equation*}
\begin{aligned}
	m_X &= \frac{c_1 - a c_2}{2a} = \frac13 \left(1 - \frac{2}{15} a^2 + \frac{2}{105} a^4  \cdots \right),\\
	m_Z &= \frac{a c_1 - 1}{4a^2} = \frac1{12} \left(1 - \frac{1}{15}a^2 + \frac{2}{315}a^4 \cdots \right),\\
\end{aligned}
\end{equation*}
where $a = \mr\dt/2$. We can easily see that $m_X \to \frac{1}{3}$ and $m_Z \to \frac{1}{12}$ as $\dt \downarrow 0$, with $O(\dt^2)$ as the leading order of truncation. Furthermore, we obtain the following asymptotic expansions of $I_\nu(z)$ and $z$:
$$ I_{\nu}(z) = \frac{e^z}{\sqrt{2\pi z}} \left(1 - \frac{4\nu^2 - 1}{8z} + \cdots\right) \qtext{and} z = \phi_\dt(\mr)\sqrt{\var_i \var_{i+1}} = \frac{2\mr/\vov^2}{\sinh\frac{\mr\dt}{2}}\sqrt{\var_i \var_{i+1}} \to \frac{4\sqrt{\var_i \var_{i+1}}}{\vov^2 \dt}.
$$
We also consider $E(\eta_{i,i+1})$ in the $h\downarrow 0$ limit:
\begin{equation*}
E(\eta_{i,i+1}) =2 \mu_i = \frac{z\,I_{\nu + 1}(z)}{2I_{\nu}(z)} \to \frac{z}{2}\left(1 - \frac{2\nu + 1}{2z}\right) = \frac{z}{2} - \frac{\delta}{4} \to \frac{2\sqrt{\var_i \var_{i+1}}}{\vov^2 \dt} - \frac{\delta}{4},
\end{equation*}
from which we obtain
\begin{equation*}
\frac{\delta}{2}+2 \mu_i = \frac{\delta}{2}+ \frac{z\,I_{\nu + 1}(z)}{I_{\nu}(z)} \to \frac{\delta}{2}+ \frac{4\sqrt{\var_i \var_{i+1}}}{\vov^2 \dt} - \frac{\delta}{2} = \frac{4\sqrt{\var_i \var_{i+1}}}{\vov^2 \dt} \qtext{as} h\downarrow 0.
\end{equation*}
Combining these results, we obtain
\begin{equation*}
	\begin{aligned}
	E\left(\hat{I}_{i,i+1}^\text{POIS}\right) &= (\var_i + \var_{i+1})m_X h + \left(\frac{\delta}{2} + 2E(\mu_i\,|\,\var_i, \var_{i+1})\right) m_Z \vov^2 \dt^2 \\
	&= (\var_i + \var_{i+1})m_X h + \left(\frac{\delta}{2} + 2E(\eta_{i,i+1})\right) m_Z \vov^2 \dt^2 \\
& \to (\var_i + \var_{i+1}) \frac{\dt}3 + \frac{4\sqrt{\var_i \var_{i+1}}}{\vov^2 \dt}  \frac{\vov^2 \dt^2}{12}
	= \left(\var_i + \var_{i+1} + \sqrt{\var_i \var_{i+1}}\right)\frac{\dt}{3}.
	\end{aligned}
\end{equation*}
This result is consistent with \citet[Proposition~3.5]{tse2013lowbias}.

\begin{singlespace}
\bibliography{Heston}
\bibliographystyle{elsarticle-harv}
\end{singlespace}

\end{document}